# LUNA: a Laboratory for Underground Nuclear Astrophysics

H Costantini[1], A Formicola[2], G Imbriani[3], M Junker[2], C Rolfs[4] and F Strieder[4]

[1] INFN Genova, Italy
[2] INFN Laboratori Nazionali del Gran Sasso, Assergi, Italy
[3] Dipartimento di Scienze Fisiche, Universitá Federico II , Napoli, and INFN Napoli, Italy
[4] Fakultät für Physik und Astronomie, Ruhr-Universität Bochum, Germany

**Abstract:** It is in the nature of astrophysics that many of the processes and objects one tries to understand are physically inaccessible. Thus, it is important that those aspects that can be studied in the laboratory be rather well understood. One such aspect are the nuclear fusion reactions, which are at the heart of nuclear astrophysics. They influence sensitively the nucleosynthesis of the elements in the earliest stages of the universe and in all the objects formed thereafter, and control the associated energy generation, neutrino luminosity, and evolution of stars. We review a new experimental approach for the study of nuclear fusion reactions based on an underground accelerator laboratory, named LUNA.

# 1. Introduction

Investigations during the last century have shown **[1-3]** that we are connected to distant space and time not only by our imagination but also through a common cosmic heritage: the chemical elements that make up our bodies. These elements and their isotopes were created by nuclear fusion reactions in the hot interiors of remote and long-vanished stars over many billions of years. Their nuclear fuels finally spent, these giant stars met death in cataclysmic explosions, called supernovae, scattering afar the atoms of heavy elements synthesised deep within their cores. Eventually this material, as well as material lost by smaller stars during the red-giant stages, collected into clouds of gas in interstellar space. These, in turn, slowly collapsed giving birth to new generations of stars, thus leading to a cycle of evolution that is still going on. In this scenario, the sun and its complement of planets were formed some 5 billion years ago. Drawing upon the material gathered from the debris of its stellar ancestors, the planet earth provided the conditions that eventually made life possible. Thus, like every object in the solar system, each living creature on earth embodies atoms from distant corners of our galaxy and from a past thousands of times more remote than the beginning of human evolution. Thus, in a sense, each of us has been inside a star and truly and literally consists of stardust. In a sense, each of us has been in the vast empty space between the stars; and - since the universe has a beginning - each of us was there. Every molecule in our bodies contains matter that once was subjected to the tremendous temperatures and pressures at the centre of a star. This is where the iron in our blood cells all originated, the oxygen we breathe, the carbon and nitrogen in our tissues, and the calcium in our bones. All were formed predominantly in fusion reactions of smaller atoms in the interior of stars. The smaller atoms themselves, i.e. hydrogen and helium, were created prior to star formation in the very early universe, the big bang.

The theories of nucleosynthesis have identified the most important sites of element formation and also the diverse nuclear processes involved in their production. The detailed understanding of the origin of the chemical elements combines astrophysics and nuclear physics, and forms what is called nuclear astrophysics. Nuclear fusion reactions are at the heart of nuclear astrophysics. They influence sensitively the nucleosynthesis of the elements in the earliest stages of the universe and in all the objects formed thereafter, and control the associated energy generation, neutrino luminosity, and evolution of stars. A good knowledge of the rates of these reactions is thus essential for understanding the broad picture outlined above.

# 2. Stellar evolution and nuclear reactions

## 2.1 The losing battle against gravity

A star like our sun is born from an interstellar cloud by gravitational contraction on a time scale typically 10 million years. However, this gravitational collapse stops after this time and the star reaches a stable situation. The reason is the following. Consider the surface layer of the sun with a temperature of 5800 K. This layer presses with its weight on the next lower layer, which must have thus a higher temperature in order that its heat pressure can balance the overlaying weight. But the surface layer and this second layer press even more on the next deeper layer, which then must have a still higher temperature. If one continues this consideration to deeper and deeper layers, one reaches eventually the centre of the sun, for which the astrophysicists have calculated a temperature of $15 \times 10^6$ K. This high temperature must be sustained by an energy source, which is derived from the H-burning process: $4\,^1H \rightarrow\,^4He + 2e^+ + 2\upsilon$ with an energy release of 26 MeV per process. In order to win the battle against the ever pressing gravitational force, the sun must consume about 700 million tons of hydrogen in every second. Taking into account the mass of the sun and the fact that only 10% of the central solar mass can reach the needed high temperature, the lifetime $\tau$ of the sun with its present luminosity L - the energy radiated into space per unit of time - is about 10 billion years. This is consistent with the observations of geology and palaeontology. More massive stars than the sun must spend even more hydrogen in the battle: one finds a proportionality $L \sim M^4$, where M is the stellar mass. Due to the larger luminosity, massive stars have a higher surface temperature than the sun and also their lifetime $\tau$ is significantly shorter than that of the sun, scaling roughly as $\tau \sim 1/M^2$. For example, a star with a 20 fold higher mass than the sun will live only for about 25 million years. A star at this stage of evolution, i.e. H-burning, is called a main-sequence star.

When a star has consumed its hydrogen in the core, the central material is replaced by the ashes $^4$He of the H-burning. These ashes can again gravitationally contract until the core reaches a sufficiently high temperature - higher than in the H-burning -, so that helium burning can proceed: $3\,^4$He → $^{12}$C and $^{12}$C$(\alpha,\gamma)^{16}$O with an energy release of about 8 MeV per reaction. In this way, a star is again stabilized winning his second battle against gravity. Due to the smaller energy release, this He-burning phase has a shorter lifetime than that of H-burning. As a result of the He-burning carbon and oxygen are produced in the innermost region of the star. After the central He exhaustion, the stellar CO core contracts and the core temperature rises. Only stars with masses greater than 10 solar mass develop temperatures large enough for further nuclear burning phases. The game between "pressing" and "heating" continues for these stars with C-burning, O-burning, Ne-burning, and Si-melting [1-7], whereby a star produces eventually elements up to the Fe region. At this stage, the nuclear matter has reached its highest binding energy per nucleon and consequently no further nuclear energy source is available. Now, gravity wins the battle and the star collapses. If the core of the star has at this stage a mass less than 1.5 solar mass, the gravitational collapse is stopped only by the electrons via the Pauli principle and the star becomes a white dwarf, the size of the earth. If the mass is below 2.5 solar mass the star becomes a neutron star, with a typical size of 20 km - again stabilized by the Pauli principle of the neutrons. Moreover, when the core reaches the nuclear density a rebounce occurs leading to an outward moving shock wave, this is a so called type II supernova explosion. If the mass is higher than 2.5 solar mass it becomes a black hole: a journey with no return.

In low and intermediate mass stars, the development of a high pressure from electron degeneracy and the cooling caused by the production of neutrinos prevent the onset of the C-burning. These stars conclude their life as CO white dwarfs. In case of a close binary system such a CO white dwarfs might accrete mass from its companion and thus explode as a nova or a type Ia supernova.

In all cases the star ejects in the evolution its outer envelope into space, e.g. as a ring-nebula or as an explosive debris, such as in supernovae.

*2.2 Nuclear reaction rate*

In an astrophysical plasma the constituent nuclei are usually in thermal equilibrium at some local temperature T. Occasionally they collide with other nuclei, whereby two different nuclei can emerge from the collision, 1+2→3+4 or A+x→B+y or A(x,y)B, i.e. a nuclear reaction. The nuclei 1 and 2 form the entrance channel of the reaction and the nuclei 3 and 4 - the ejectiles - form the exit channel of the reaction. If the nuclear reaction Q-value, $Q = (m_1+m_2-m_3-m_4)c^2$ ($m_i$ = nuclear masses, c = velocity of light), is positive, there is a net production of energy in the reaction for each event. Such reactions are most important for the energy generation in stars. Of equal importance is the intrinsic probability that a given reaction will take place. This probability, expressed as an energy-dependent cross section $\sigma(E)$, determines how many reactions occur per unit time and unit volume. Hence, together with the Q-value, $\sigma(E)$ provides important information on nuclear energy generation in stars.

The cross section $\sigma(E)$ of a nuclear fusion reaction is of course governed by the laws of quantum mechanics where, in most cases, the Coulomb and centrifugal barriers arising from nuclear charges and angular momenta in the entrance channel strongly inhibit the penetration of one nucleus into another. This barrier penetration leads to a steep energy dependence of the cross section. Other energy-dependent effects, such as resonances and their mutual interference effects, also play important roles and the energy dependence of $\sigma(E)$ can be quite complex. It is the challenge of the experimentalist to make precise $\sigma(E)$ measurements over a wide range of energies, as our fragmented knowledge of nuclear physics prevents us from predicting $\sigma(E)$ with sufficient precision on purely theoretical grounds. This challenge represents the major goal of this review.

In the hot stellar matter the energies of the moving nuclei can be described usually by a Maxwell-Boltzmann distribution, $\Phi(E) \propto E \exp(-E/kT)$, where k is the Boltzmann constant. Folding the cross section with this energy (or velocity) distribution leads to the nuclear reaction rate per pair of nuclei [3]

$$\langle\sigma v\rangle = (8/\pi\mu)^{1/2} (kT)^{-3/2} \int_0^\infty \sigma(E) \exp(-E/kT)\, dE, \qquad (1)$$

where v is the relative velocity of the pair of nuclei, E is the centre-of-mass energy, and $\mu = m_1 m_2/(m_1+m_2)$ is the reduced mass of the entrance channel. We present in this review beam energies always in the centre-of-mass system, except where quoted differently. In order to cover the different evolutionary phases of stars, i.e. from main-sequence stars such as our sun (T ≈ $10^7$ K) to supernovae or the big bang (T ≈ $10^9$ K), one must know the reaction rates over a wide range of temperatures, which in turn requires the availability of σ(E) data over a wide range of energies.

For the important class of charged-particle-induced fusion reactions, there is a repulsive Coulomb barrier in the entrance channel of height $E_c = Z_1 Z_2 e^2/r$, where $Z_1$ and $Z_2$ are the integral nuclear charges of the interacting particles, e is the unit of electric charge, and r is the radius. Due to the tunneling effect through the Coulomb barrier, the cross section drops nearly exponentially with decreasing energy (figure 1):

$$\sigma(E) = S(E)\, E^{-1} \exp(-2\pi\eta), \qquad (2)$$

where $\eta = 2\pi Z_1 Z_2 e^2/hv$ is the Sommerfeld parameter (h = Planck constant). The function S(E) - defined by this equation - contains all nuclear effects and is referred to as the nuclear or astrophysical S(E) factor. If equation (2) is inserted in equation (1), one obtains

$$\langle\sigma v\rangle = (8/\pi\mu)^{1/2} (kT)^{-3/2} \int_0^\infty S(E) \exp(-E/kT - b/E^{1/2})\, dE, \qquad (3)$$

with $b = 2(2\mu)^{1/2} \pi^2 e^2 Z_1 Z_2 / h$. Since for nonresonant reactions the S(E) factor varies slowly with energy (figure 1), the steep energy dependence of the integrand in equation (3) is governed primarily by the exponential term, which is characterised by a peak at an energy $E_0$ that is usually much larger than kT, the mean thermal energy in the stellar plasma. The peak is referred to as the Gamow peak (figure 1); for a constant S(E) value over the energy region of the peak, one finds $E_0 = (bkT/2)^{2/3}$, the Gamow energy. This is the effective mean energy for a given reaction at a given temperature. Approximating the peak by a Gaussian function, one finds an effective width $\Delta = 4(E_0 kT)^{1/2}/3^{1/2}$. Thus, nuclear burning takes place predominantly over the energy window $E_0 \pm \Delta/2$, the thermonuclear energy range, for which information on the cross section σ(E) must be obtained. Due to the high sensitivity of the reaction rate on the height of the Coulomb barrier, there are the very distinct burning phases such as H-, He-, and C-burning. Even if a star has available equal amounts of H, He and C materials, the star burns first H, while the burning of He and C is completely negligible at this stage.

*2.3 Extrapolation of laboratory data*

Due to the steep drop of the fusion cross section at subcoulomb energies, it becomes increasingly difficult to measure σ(E) as E is lowered. Although experimental techniques have improved significantly over the years [3,4], extending σ(E) measurements to lower and lower energies - with a low-energy limit at $E_L$, corresponding to a reaction yield in a detector of roughly 1 event per hour -, it has not been possible to measure σ(E) at thermonuclear energy. Instead, the measured energy dependence of σ(E) at higher energies must be extrapolated to stellar energies, using the S(E) factor defined in equation (2). Such an "extrapolation into the unknown" can lead to considerable uncertainty. At energies lower than $E_L$ there might be a change of reaction mechanism, or of the centrifugal barrier, or there might be a contribution of narrow or subthreshold resonances to σ(E). The danger of such extrapolations was strikingly demonstrated [8] in the case of the reaction $^2H(d,\gamma)^4He$, involved in the big-bang nucleosynthesis. The new low-energy data increased the extrapolated values by a factor 1000, mainly due to a change in the centrifugal barrier of the entrance channel, i.e. the orbital angular momentum between the deuterons changed from d-partial waves to s-partial waves. In turn, the new data are a clear demonstration of the existence of a D-state admixture in the wavefunction of the $^4$He ground state, of the order of a few percent. Thus, the strongly bound double-magic nucleus $^4$He is not spherical in shape, as might be expected, but has an intrinsic quadrupole moment. The few percent D-state admixture in the wave function of the $^4$He ground state is responsible for the factor 1000 increase in cross section at low energies: a fine tuning in nuclear physics. Note that these low-energy studies appeared at first - from a purely nuclear physics point of view - to be of comparatively little interest. However, as shown by this example and many others, acquisition and evaluation of improved data often provide unexpected intellectual rewards in nuclear physics.

Yet another example is the reaction $^{12}C(\alpha,\gamma)^{16}O$, which takes place in the helium-burning phase of red-giant stars producing predominantly carbon and oxygen. We still cannot show in the laboratory and in theoretical calculations why the ratio of oxygen to carbon in the sun and similar stars is close to two-to-one. We humans are mostly (84% by mass) oxygen and carbon. We understand in a general way the chemistry and biology involved, but we certainly do not understand the nuclear astrophysics which produced the oxygen and carbon in our bodies. It has been found theoretically that the above reaction influences sensitively not only the nucleosynthesis of the elements between carbon and iron but also the evolution of stars, the dynamics of supernovae, and the kind of remnant - neutron star or black hole - of a supernova explosion. According to stellar model calculations, the cross section $\sigma(E_0)$ of $^{12}C(\alpha,\gamma)^{16}O$ at the relevant Gamow energy ($E_0$ = 0.3 MeV) should be known with a precision of at least 10% [7,11]. In spite of tremendous experimental efforts over nearly 35 years ([6,12] and references therein), one is still far from this goal.

Thus, low-energy, high intensity accelerator and in some cases new experimental approaches are needed to reduce the uncertainties of all these $\sigma(E)$ extrapolations. However, as we will demonstrate below measurements at earth's surface are hampered by cosmic-ray induced background. It is likely that significant progress for direct measurements at or very close to the Gamow window for most of the astrophysical relevant reactions can only be reached with accelerator facilities in underground laboratory.

*2.4 Hydrogen-burning and related solar physics*

Most of this review is focused on reactions that occur during the H-burning, because with LUNA mainly these reactions could be studied using proton, $^3$He, and $^4$He beams but not a deuterium beam (forbidden at LUNA). This burning phase is ignited when the centre of a star reaches a temperature of about $6\times10^6$ K and the proton-proton reaction can synthesize deuterium. In the hydrogen rich environment the deuterium reacts almost instantly with another proton producing $^3$He, which in turn forms – together with another $^3$He nucleus – an alpha particle and releases two protons: $^3$He($^3$He,2p)$^4$He. This represents the main termination of the pp-chain with a net conversion $4p \to {}^4$He + $2e^+ + 2\nu_e$ + 26.73 MeV. The channels $^3$He+p and $^3$He+d are excluded or very rare due to the unbound $^{4,5}$Li nuclei and the low deuterium abundance, respectively. The other branch of the pp-chain proceeds through the reaction $^3$He($^4$He,$\gamma$)$^7$Be with an additional branching through the electron capture of $^7$Be or a subsequent proton capture followed by the beta delayed alpha decay of $^8$B, resulting in two $^4$He nuclei. The final result is always the transformation of four protons into a helium nucleus accompanied with the emission of neutrinos (figure 2a).

In the case of massive stars, the core temperature reaches values larger than $15 \cdot 10^6$ K. Thus, the carbon present in the core of the star can react efficiently with the proton sea, producing $^{13}$N. This reaction starts the CNO-cycle. After a series of proton captures and decays four protons are transformed again into a helium nucleus with the emission of two neutrinos (figure 2b). At this temperature the energy rate generated by the CNO-cycle is larger than that by the pp-chain; moreover its temperature dependence is much steeper. The sum of the initial carbon and nitrogen abundances remains constant through the cycle. The $^{14}$N(p,$\gamma$)$^{15}$O has the lowest cross section among the reactions in the cycle, therefore at the end of this H-burning phase, almost all carbon is transformed into nitrogen.

In conclusion, stars that have a mass equal to or lower than that of the sun burn hydrogen via the pp-chain and live several billion years. In comparison, more massive stars can reach the temperatures needed to ignite the CNO-cycle and their life span as a Main Sequence star is much shorter than for low-mass stars, i.e. about several millions of years.

In both cases independent of the main hydrogen-burning sequence the neutrinos carry the information directly from the interior of the star. The observation of these neutrinos from the Sun in the Homestake gold mine, South Dakota, USA [16, 17, 18], was one of the scientific highlights of the last century. For the first time there was direct evidence for the hypothesis of nuclear energy generation in stars, as suggested in the pioneering work by Bethe and von Weizsäcker [139, 140]. However, the observed solar neutrino fluxes in the existing neutrino detectors were - for many years - not consistent with the standard picture of the microscopic processes in the sun: the standard solar model **[9]**. Consequently, the missing solar neutrinos were for the next three decades one of the most intriguing problems in physics and astrophysics.

A possible solution for this "solar neutrino problem" was suggested to be found in one or more of the areas of neutrino physics, solar physics, or nuclear physics. In view of the important and fundamental conclusions, which may be deduced from the results of these neutrino observations, it is of utmost importance to place all predictions on a solid basis. Thus, the standard solar model was continuously improved and verified. These improvements included the determination of the interior sound speed by helioseismology [141] and additional solar neutrino measurements from new experiments like SAGE [142], GALLEX/GNO [143, 144], Super Kamiokande [145]. All observations confirmed the need for physics beyond the standard model of electroweak theory. Finally, the simultaneous observation of all 3 neutrino families – the neutrino flavours $\nu_e$, $\nu_\mu$, and $\nu_\tau$ – in the Sudbury Neutrino Observatory (SNO) [10] has not only given convincing evidence for solar neutrino oscillations, but also confirmed the neutrino-flux predictions of the standard solar model proved this new physics aspects.

The LUNA experiment is an important part of this successful story of solar neutrinos. The nuclear physics component of the standard solar model involves the cross sections of the hydrogen-burning. These nuclear reaction cross sections were greatly improved over the years and excluded a nuclear solution for the missing neutrinos.

The highlight of the LUNA experimental programme was certainly a study of the reaction $^3$He($^3$He,2p)$^4$He. The solar Gamow peak in this reaction at $E_0 = 21$ keV has been reached with a detected reaction rate of about 2 events per month at the lowest energy. Thus, the cross section of an important fusion reaction of the H-burning pp-chain has been directly measured for the first time at solar thermonuclear energies. This measurement excluded a nuclear solution of the solar neutrino problem due to a resonance in the reaction $^3$He($^3$He,2p)$^4$He. The studies with the first LUNA underground accelerator were concluded with a measurement of the d(p,$\gamma$)$^3$He reaction. The first case of a capture reaction studied over the full energy range of the solar Gamow peak.

The excellent agreement between predicted and observed solar neutrino-flux implies that the experimentalists have correctly determined the rates of the nuclear reactions occurring in the Sun. However, this success does not mean that a better knowledge of the nuclear reaction rates in the sun is no longer needed. In contrast, now the aim must be to turn the sun into a calibrated neutrino source to deduce information about the neutrino masses and mixing parameters from the observations of the various neutrino detectors. Recently such Information became available from the KamLAND reactor neutrino experiment [146] and BOREXINO [83, 84].

The progress in solar neutrino experiments will allow an independent determination of the input physics of the standard solar model, such as the core temperature or the primordial abundances of carbon, oxygen and heavier elements. These parameters can be derived from a global analysis of experimental quantities like the solar neutrino-fluxes, nuclear cross sections and oscillation parameters measured in the laboratory. Moreover, the recent downward revision of the metal content of the solar convective zone [147] adds interest on a check of the standard solar model. The new metal abundances significantly alter the agreement between the standard model and helioseismology in the temperature region below the solar convective zone.

These problems demand further improved nuclear fusion data, an aim of LUNA. This aim has been already partly achieved after the installation of a new accelerator which opened the possibility to access key reactions of the hydrogen-burning important for these current research topics: the reactions $^3$He($\alpha$,$\gamma$)$^7$Be (pp-chain) and $^{14}$N(p,$\gamma$)$^{15}$N (CNO cycle). The lower-energy limit for direct measurements of these reactions was strongly decreased and the precision of the reaction data could be improved significantly.

The different nuclear reaction studied at LUNA will be described in detail in the sections 3 (LUNAI) and 4 (LUNAII). We will also discuss again the specific astrophysical relevance of a given reaction at the time when we discuss the associated exüperimental results.

## 3. The LUNA project

### 3.1 Underground laboratories

All rare-event experiments are limited by background signals due to cosmic-rays, and natural or induced radioactivity: proton decay [13], neutrinoless double $\beta\beta$ decay [14], and dark matter search [15], including low-energy studies of reactions in nuclear astrophysics induced by charged particles. A unique solution for these highly sensitive experiments is to carry them out in deep underground laboratories. The first experiment installed in such an underground laboratory was the famous neutrino detector of Ray Davis in the Homestake gold mine, South Dakota, USA [18]. The history of the Laboratori Nazionali del Gran Sasso (LNGS) in Italy (figure 3) began in 1982 with the approval by the Italian parliament. The excavation of the experimental halls, which was completed in 1987, was a by-product of the construction of a road tunnel for the highway connection between Rome and the Adriatic Sea crossing the Abruzzian Mountains. The Gran Sasso Mountain with an average thickness of 1400 m of rock or 3800 m.w.e.[1] above the laboratory halls led to a reduction of the muon flux, the most penetrating component of the cosmic-rays, by a factor $10^6$ compared with the earth's surface. It offers thus a low-background environment [19,20,21,22]. The first experiment hosted in the laboratory, MACRO [23,24], started data taking in 1989.

Thermonuclear reactions induced by charged particles are mainly studied by means of γ-ray or particle spectroscopy both hampered predominantly by background effects of natural radioactivity and cosmic-rays in the detectors. This leads typically to a background signal rate much higher than the expected count rate at ultra-low energies. The main sources of natural γ-ray background arise from several sources. (i) Radionuclides belonging to natural radioactive series ($^{226}$Ra, $^{214}$Bi, $^{214}$Pb from the $^{238}$U chain; $^{228}$Ac, $^{224}$Ra, $^{208}$Tl, $^{212}$Pb from the $^{232}$Th chain). (ii) Radon ($^{222}$Rn), a short-lived radioactive gas and a daughter product of the uranium and thorium decay chains. (iii) Long-lived natural radionuclides such as $^{40}$K, $^{87}$Rb, $^{115}$In, $^{133}$La, $^{142}$Ce, etc. (iv) Radionuclides of cosmogenic origin such as $^{3}$H, $^{7}$Be, $^{14}$C, $^{22}$Na, $^{26}$Al, $^{60}$Co. (v) Radionuclides of artificial origin, the long-lived fission products such as $^{95}$Nb, $^{95}$Zr, $^{144}$Ce, $^{106}$Ru, $^{134}$Cs, $^{125}$Sb, $^{137}$Cs. All these radionuclides lead to γ-ray background with signals below $E_\gamma = 3.5$ MeV (for an extended discussion of environmental γ-ray background, see [25, 26]). In comparison, the γ-ray background above $E_\gamma = 2.6$ MeV is mainly dominated by the effects of cosmic-rays in the detectors, i.e. muon and neutron-induced events. Conventional passive or active shielding around the detectors can only partially reduce the problem. The best solution is to install an accelerator facility in a laboratory deep underground, similar to the solar neutrino detectors and other rare event experiments. The rock shielding of LNGS leads in a HPGe γ-ray detector to a thousand-fold reduction in the γ-ray background signal above $E_\gamma = 2.6$ MeV (figure 4a), i.e. the region above natural radio-activity. In particular, in the region between $E_\gamma = 7$ and 12 MeV, the suppression factor is 100 times better of what was achieved in laboratories at the earth's surface using active muon shielding in combination with paraffin and lead [27, 28]. However, γγ-coincidence methods [28] could lead to a significantly higher suppression factor at the earth's surface compared to standard active and passive shielding, but these methods are always coupled to particular nuclear reactions and lead to a reduced detection efficiency.

Therefore, the advantage of an underground environment is evident for high Q-value reactions such as d(p,γ)$^3$He and $^{14}$N(p,γ)$^{15}$O, but appears less evident at first sight for low Q-value reactions. In a surface laboratory passive shielding such as lead can be placed around the detectors but it is limited to a certain thickness. One cannot add further shielding material since cosmic-ray muons interact always with the material and create energetic neutrons which, in turn, create γ-rays in the lead. Clearly, this background component is dramatically reduced with the significantly suppressed muon flux in an underground laboratory. An additional γ-ray background component below $E_\gamma = 2.6$ MeV arises due to neutron-induced events, where the neutrons are created by (α,n) reactions in the surrounding rocks, where the α's come from natural α-radioactivity. A separate treatment is needed for $^{222}$Rn, which diffuses out of the rocks and permeates through the detector shielding. The decaying radon and its daughters produce α and β particles that produce again secondary γ radiation by bremsstrahlung and nuclear reactions. A popular solution of this problem is to house the detector in a box with a small overpressure of flushing nitrogen (figure 4b) [29].

*3.2 The LUNA 50 kV accelerator*

As a pilot project for an underground accelerator facility, a home-made 50 kV accelerator (figure 5) was installed in the LNGS underground laboratory. This unique project, called LUNA I (LUNA = **L**aboratory for **U**nderground **N**uclear **A**strophysics), was initiated in the year 1990 by G. Fiorentini and one of the authors (C.R.) and started measurements in January 1994. Technical details of the LUNA accelerator setup have been reported elsewhere [30].

Briefly, the facility consisted of a duoplasmatron ion source mounted on a platform biased with a high voltage of up to 50 kV. The beam was extracted from the source by an extraction/acceleration system and transported to the target by a double-focusing 90° analyzing magnet. Exceptional features of this system were the small beam energy spread of less than 20 eV, an acceleration voltage known with an accuracy of better than $10^{-4}$, and a $^3$He beam current of about 300 μA even at low energies. The ion beam was injected into a windowless (recirculating) $^3$He gas target system which allowed for a pressure of 0.5 mbar in the interaction chamber. The beam was dumped on a beam stop located inside the interaction chamber which, as an effect of the beam power, heats to high temperatures. This feature was used to derive the beam current by means of a beam calorimeter, to a precision of 3%. The calorimeter method was needed since an ion beam passing through a gas target creates a plasma leading to unreliable ion beam current–readings. While the system was primarily developed for the measurements of the $^3$He($^3$He,2p)$^4$He reaction (section 3.3) it was employed successfully also for measurements of the electron screening effects (section 3.4) and the cross section of the D(p,γ)$^3$He reaction (section 3.5). The 50 kV accelerator was decommissioned in 2003. The acceleration system is now on display in the "Museum of Physics and Astrophysics" by INFN in Teramo (Italy) while gas target and calorimeter system have been refined and are now used at the LUNA 400 kV accelerator (section 4).

---

[1] m.w.e. ≡ Meter water equivalent

*3.3 The $^3$He($^3$He,2p)$^4$He reaction*

LUNA I was designed [30] for a renewed study of the reaction $^3$He($^3$He,2p)$^4$He, aiming to reach its solar Gamow peak at $E_0 \pm \Delta/2 = 21 \pm 5$ keV, while previous studies **[31]** had to stop at the high energy edge of the Gamow peak. This experimental goal has been reached [32, 33] with a detected reaction rate of about 2 events per month at the lowest energy, E = 16.5 keV, with σ = 20 fb or $2 \times 10^{-38}$ cm$^2$ (figure 1). The measurements were carried out in three steps using two different detector configurations [32, 33].
The first step was performed between E = 45 and 92 keV at the 450 kV SAMES accelerator of the Dynamitron Tandem Laboratory at the Ruhr-Universität Bochum to obtain an overlap with previous data [31]. The detector setup consisted of four ΔE-E particle detectors with a surface of 5.5 cm$^2$ each placed in the gas target at opposite sides of the beam axis. The thickness (ΔE: 140 μm, E: 1000μm) of these detector telescopes was chosen to achieve a sufficient separation between the protons emitted by the reaction $^3$He($^3$He,2p)$^4$He and a proton background induced by the reaction $^3$He(d,p)$^4$He (figure 6). Deuterium is a nearly unavoidable source of beam-induced background and was present mainly in the beam (as a HD$^+$ molecule with mass 3) as well as a minor contamination in the target gas. During the experiment the D/$^3$He ratio was between $5 \times 10^{-6}$ and $1 \times 10^{-7}$. At a beam energy of $E_{lab}$ = 40 keV the cross section of $^3$He(d,p)$^4$He is 6 orders of magnitudes higher than that $^3$He($^3$He,2p)$^4$He, mainly due to the charge dependence of the penetration probability through the Coulomb barrier. Therefore, small overlaps in the regions of interest for the signals from both reactions could cause a significant background for the evaluation of the $^3$He($^3$He,2p)$^4$He cross section. Indeed, about 0.4% of the protons from $^3$He(d,p)$^4$He create events in the energy region of the protons from $^3$He($^3$He,2p)$^4$He. Thus, a minimization of the deuterium contamination to a minimum was needed and achieved. The resulting data were in perfect agreement with previous data [31] (figure 1). However, the cosmic-ray background limited the measurements above the earth's surface. The large active detector areas and the close geometry of the two detectors of the ΔE-E telescopes led to a high sensitivity of this detection system for muon-induced events: about 10 events/day per detector in the region of interest.
Subsequently, the gas target and detection setup have been moved to LNGS, together with the home-made 50 kV accelerator. The background induced by cosmic-rays in the experiment was reduced now from 10 events/day at earth's surface to less than 1 event/month in the underground laboratory (figure 7). However, a similar beam-induced background as before remained: the protons from $^3$He(d,p)$^4$He limited the studies with this setup to energies above E = 20.7 keV [30]. To overcome this limitation a new detector setup was placed around the beam axis forming a 12 cm long parallelepiped box in the target chamber. This setup consisted of eight detectors with an area of 5.5 cm$^2$ each. The cross section was evaluated applying several coincidence conditions [34] between the detectors in order to focus on the pp coincidences: an event had to contain hits in two (and not more than two) different detectors. In this way the natural background as well as beam-induced background could be reduced significantly. The reliability has been checked during a 23-day background run with a $^4$He beam: no event has been detected fulfilling the selection criteria and, therefore, excluding any contribution from the $^3$He(d,p)$^4$He reaction. This setup and procedure allowed to measure the cross section down to E = 16.5 keV. At this energy the reaction cross section is σ = 20 ± 20 fb corresponding to a rate of 2 events/month [34] (figure 1). Thus, the cross section of an important fusion reaction of the H-burning pp-chain has been directly measured for the first time at solar thermonuclear energies; an extrapolation was no longer needed in this reaction: the dream of Willy Fowler became reality. In the 1970's it was suggested that a low-energy narrow resonance within the solar Gamow peak could exist in this reaction, which could solve or minimize the solar neutrino problem [35, 36]. The LUNA data could not find such a resonance and consequently a nuclear physics solution of the solar neutrino problem was definitely excluded.

*3.4 Electron screening and Stopping Power at low energies*

An important aspect of ion-beam experiments at the ultra-low energies is demonstrated in the measurement of the $^3$He($^3$He,2p)$^4$He reaction: the electron screening effect (figure 1). This effect was already known before from experiments at earth's surface [38, 39] and has a serious impact for the transformation of direct measurements at the relevant energies into an astrophysical reaction rate. In the extrapolation of the cross section using equation (2), it is assumed that the Coulomb potential of the target nucleus and projectile is that resulting from bare nuclei. However, for nuclear reactions studied in the laboratory, the target nuclei and the projectiles are usually in the form of neutral atoms or molecules and ions, respectively. The electron clouds surrounding the interacting nuclides act as a screening potential: the projectile effectively sees a reduced Coulomb barrier, both in height and radial extension. This, in turn, leads to a higher cross section for the screened nuclei, $\sigma_s(E)$, than would be the case for bare nuclei, $\sigma_b(E)$. There is, in fact, an enhancement factor [38],

$$f_{lab}(E) = \sigma_s(E)/\sigma_b(E) \approx \exp(\pi\eta U_e/E) \geq 1 ,  \qquad (4)$$

where $U_e$ is an electron-screening potential energy. This potential energy can be calculated, for example, from the difference in atomic binding energies between the compound atom and the projectile plus target atoms of the entrance channel, or from the acceleration of the projectiles by the atomic electron cloud. Note that for a stellar plasma, the value of the bare cross section $\sigma_b(E)$ must be known because the screening in the plasma could be quite different from that in the laboratory nuclear-reaction studies, i.e. $\sigma_p(E) = f_p(E) \sigma_b(E)$, where the plasma enhancement factor $f_p(E)$ must be explicitly included for each situation. A good understanding of electron-screening effects in the laboratory is needed to arrive at reliable $\sigma_b(E)$ data at low energies. An improved understanding of laboratory electron screening may also help eventually to improve the corresponding understanding of electron screening in stellar plasmas, such as in our sun.

Recent experimental studies of reactions involving light nuclides ([40] and references therein) – all carried out at surface laboratories – have verified the expected exponential enhancement of the cross section at low energies. The LUNA collaboration contributed to all these studies with a single experiment: the measurement of the d($^3$He,p)$^4$He cross section with a low-energy limit at $E_L = 4.2$ keV [41]. However, the observed enhancements were in several cases larger (about a factor 2) than could be accounted for from available atomic-physics models, i.e. the adiabatic limit $U_{ad}$. It should be noted that the studies of the electron screening effect can be carried out on reactions not necessarily relevant to astrophysics. These studies often do not require an underground laboratory.

The measurement of the d($^3$He,p)$^4$He reaction in inverse kinematics at the Gran Sasso underground laboratory (figure 8) was performed to determine not only the magnitude of the screening effect in this reaction, but also to establish values for the energy loss used in the data reduction. Due to the steep drop of the cross section at subcoulomb energies (figure 1), an accurate knowledge of the effective beam energy associated with the observed reaction yield is as important as the yield measurements themselves, a general rule. In the analysis of such data, the effective beam energy in the target involves always energy-loss corrections, which are extracted from a standard stopping-power compilation [42]. The observed stopping power of the $^3$He ions in the D$_2$ gas target was in good agreement with the extrapolation from standard compilations [42]. Using these stopping power values the data led to an electron-screening potential for the d($^3$He,p)$^4$He reaction of $U_e = 132 \pm 9$ eV, which is significantly higher than the estimated adiabatic limit (including a Coulomb-explosion process of the D$_2$ target molecules) of $U_{ad} = 65$ eV. In conclusion, the lack of information on reliable low energy stopping power data is unlikely to be responsible for the discrepancy between experiments and atomic models. However, that might not always be the case. The compilation is based on experimental data down to energies around the Bragg peak (i.e. the maximum in the stopping-power curve), while at lower energies - relevant to nuclear astrophysics - the experimental data are often extrapolated with theoretical guidance. For example, for the studies of the reaction $^3$He(d,p)$^4$He stopping-power data were needed for deuterons in $^3$He gas at energies $E_d$ below 100 keV, i.e. below the region of the Bragg peak. For the measurements an "astrophysical" method was used [43, 44]. The deduced stopping-power values as a function of $E_d$ showed a surprising threshold behaviour near $E_d = 18$ keV (figure 9). In a simple model, the threshold in the electronic stopping power arises from the minimum energy transfer $E_{e,min} = 19.8$ eV in the 1s→2s electron excitation of the He target atoms, which translates into a minimum deuteron energy $E_{d,min} = 18.2$ keV. Below this deuteron energy, the electron cloud of the He atom cannot be excited via an ion-electron interaction and thus the electronic energy loss vanishes: the He atoms become essentially transparent for the deuterons. Below the threshold energy, solely the nuclear stopping power is left. The observed threshold behaviour is a quantum effect and may be compared in a way with electron superconductivity or the K-edge in X-ray absorption. Thus, stopping-power data at energies below the Bragg peak may have to be obtained in low-energy studies of astrophysical reactions, which however may not need an underground laboratory.

Recently, the electron screening in d(d,p)t has been studied for deuterated metals, insulators, and semiconductors in surface experiments ([45] and references therein). As compared to measurements performed with a gaseous $D_2$ target ($U_e = 25\pm5$ eV, $U_{ad} = 27$ eV), a large screening was observed in all metals (of order $U_e = 300$ eV), while a small (gaseous) screening was found for the insulators and semiconductors. Unfortunately, experiments with a deuterium beam cannot be performed at LNGS due to a possible neutron production, which in turn could create background signals in other low-level experiments at LNGS. An expected correlation of the electron screening effect with the nuclear charge $Z_t$ of the target atoms was verified ([46] and references therein) in the reactions $^7$Li(p,α)α and $^6$Li(p,α)$^3$He, $^9$Be(p,α)$^6$Li and $^9$Be(p,d)$^8$Be, $^{50}$V(p,n)$^{50}$Cr, and $^{176}$Lu(p,n)$^{176}$Hf, always for pure metals and alloys. The data demonstrate that the enhanced electron screening occurs across the periodic table and is not restricted to reactions among light nuclides. The two reactions with neutrons in the exit channel demonstrate furthermore that the electron screening is an effect in the entrance channel of the reaction and not influenced by the ejectiles of the exit channel. An improved theory is highly desirable to explain why this large screening effect appears.

*3.5 The d(p,γ)$^3$He reaction*

The weak-interaction process p(p,e$^+$n)d effectively controls the rate of the pp-chain, i.e. the energy generation in stars like our sun. The created deuterium is quickly converted into $^3$He by the d(p,γ)$^3$He reaction (Q = 5.493 MeV). Therefore, the latter has no influence on energy generation and nucleosynthesis in H-burning. However, during the pre-main sequence phase of a star well before the onset of H-burning, an important d-burning phase takes place [47]. The main source of energy of a proto-star - formed by accretion of interstellar material onto a small contracting core - is the gravitational force. When the stellar core temperature due to the contraction reaches $10^6$ K, the primordial deuterium (a mass fraction of about $2\times10^{-5}$) is processed while the p+p fusion rate is still low and H-burning cannot take place. The main effect of the onset of d-burning is to slow down the contraction and thus the heating of the stellar core. Consequently, the lifetime of the proto-star increases and its observational properties (surface luminosity and temperature) are frozen until the primordial deuterium is fully consumed. This d-burning phase is of particular importance because, due to the slow evolutionary time-scale, a large fraction of the proto-stars is observed during this evolutionary phase. Indeed, in this proto-stellar phase the deuterium consuming reaction d(p,γ)$^3$He has a strong influence on the time scale of the d-burning due to its large cross section and the large amount of available protons. The relevant energy range for the d(p,γ)$^3$He reaction during d-burning in proto-stars is $E_0 = 1$ to 2 keV.

Furthermore, the d(p,γ)$^3$He reaction is of critical importance for cosmology: during the era of big-bang nucleosynthesis (BBN) the reaction is in competition with the universal expansion, by which the proton density is reduced. The d(p,γ)$^3$He reaction is one of the leading processes responsible for the destruction of deuterium and the concurrent $^3$He synthesis at energies between $E_0 = 25$ and 120 keV. Nevertheless, when the deuterium formation channel is opening, this role is played instead by the strong reaction d(d,n)$^3$He [48].

The d(p,γ)³He reaction proceeds through a direct capture process with the emission of a 5.5 MeV γ-ray. The reaction has been investigated previously by two groups [49, 50] (figure 10). Both data sets could not cover the energies below E = 10 keV, i.e. the solar Gamow peak and the d-burning energy region, and exhibit a 40% systematic discrepancy in the extrapolation to zero energy. The measurement at LUNA [51] was carried out using a deuterium gas target coupled with a BGO detector of large solid angle. The detection efficiency for the 5.5 MeV γ-rays was about 70% with an energy resolution of 8% [52]. Since the γ-ray background due to natural radioactivity stops around $E_\gamma$ = 3 MeV (section 3.1) the detection of the capture γ-ray transition could take full advantage of an environment almost free of cosmic-ray induced γ-ray background. The LUNA data covered the full energy range of the solar Gamow peak [51], i.e. as low as E = 2.5 keV, confirming the previous data [49] and firmly establishing the low-energy behavior of the astrophysical S factor important for the proto-star evolution and BBN. The experiment represented the first case of a capture reaction studied underground. Furthermore, the data allowed a direct test of calculations based on three-body electromagnetic currents. Indeed, the S-factor calculated by [53] is in perfect agreement with the LUNA results and proved the fundamental role of non-nucleonic degrees of freedom in the cross section determination of the d+p radiative capture reaction.

In summary, the work discussed so far demonstrated the research potential of LUNA and that all of the experimental requirements in such low-rate and time-consuming experiments can be fulfilled.

**4. LUNA II**

*4.1 The 400 kV accelerator*

With the scientific success of LUNA I the financial support improved significantly. Thus, in 2000, a commercial high-current 400 kV accelerator from High Voltage Engineering Europe could be installed at LNGS - called LUNA II (figure 11) -, which opened the possibility of improving our knowledge for other key reactions in nuclear astrophysics. The electrostatic accelerator is imbedded in a tank, which is filled with a gas mixture $N_2/CO_2$ at 20 bar. The high voltage (HV) is generated by an Inline-Cockcroft-Walton power supply (located inside the tank). The system is expected to have a HV-ripple of 30 Vpp and a reproducibility as well as long-term stability of the HV of 20 Vpp at 400 kV over several days. These parameters are consistent with observation [54]. The radio-frequency ion source - mounted directly on the accelerator tube - provides ion beams of 1 mA hydrogen (75%H$^+$) and 500 µA He$^+$ over a continuous operating time of about 40 days. The ions are extracted by an electrode, which is part of the accelerator tube. With a first 45° magnet (30 cm radius, 3 cm gap, 1.6 MeV amu; 1×10$^{-4}$ stability/h) and a vertical steerer located before the magnet, the ion beam is guided and focused properly to a gas target station. Alternatively, the ion beam can be guided through the 0° port of the magnet and a second 45° magnet (identical to the first magnet) into an additional beam line with a solid target station. The latter was installed in 2006 and allows now to prepare two different experiments. In the energy range 150 to 400 keV, the accelerator provides a proton beam current on target of up to 500 µA and a He-beam current of up to 250 µA, with a half-angle divergence of 0.3°; at 50 keV, the proton current is about 150 µA [54]. The accelerator is controlled by a PLC-based computer, which allows for a safe operation over long periods of running time without the constant presence of an operator on site.

The absolute energy $E_B$ and the energy spread $\Delta E_B$ of the proton beam have been measured at (p,γ)-resonances of ²⁵Mg, ²⁶Mg, and ²³Na in the energy range between 300 and 400 keV. The parameters of these resonances are known to high accuracy. Furthermore, the ¹²C(p,γ)¹³N non-resonant reaction has been used as an alternative method to determine the absolute proton energy over a wide energy range. An advantage in using carbon as target was that the results are not influenced by any C-deposition on the target during the runs as this might be the case for other reaction studies. In this way, the calibration error could be kept below 50 eV and the $\Delta E_B$ values obtained for the different resonances led to an upper limit of $\Delta E_B \leq 100$ eV [54]. The energy calibration of the γ-ray spectra has been checked at various times and was found to be stable within ±100 eV over several weeks. The precise determination of the beam energy as well as a small energy spread is of utmost importance for measurements at ultra-low energies, i.e. for $E_B$ < 100 keV. Let us take the ¹⁴N(p,γ)¹⁵O reaction: due to the nearly exponential drop of the cross section σ(E) with decreasing energy, an error of 1.5 keV in beam energy at $E_p$ = 100 keV leads to an error in $\sigma(E)$ of about 20%. The presently measured uncertainty of the beam parameters amounts to a 5% uncertainty in $\sigma(E)$ at this energy. For similar reasons, the absolute energy spread and long-term energy stability must be known sufficiently well, where the latter feature is particularly important in view of the long running times (several days) at low energies.

*4.2 The $^3$He($\alpha,\gamma$)$^7$Be reaction*

The $^3$He($\alpha,\gamma$)$^7$Be reaction represents presently the largest uncertainty in the prediction of the flux of solar neutrinos and was considered in the past as a possible key to solve the solar neutrino puzzle. The successful experiments of SNO and Kamland [10, 37] proved the existence of neutrino oscillations and gave an explanation of the observed solar neutrino deficit in earth neutrino detectors. This success opened a new era of neutrino spectroscopy, in which the solar neutrino fluxes serve as a probe for details of the standard model of particle physics. In particular, the precise knowledge of the different neutrino fluxes (figure 2) can be used to understand chemical properties of the sun, provided that nuclear reaction cross sections are known with similar accuracy. It appears possible to exploit neutrinos from the CNO-cycle and pp-chain to determine the primordial solar core abundances of C and N, and in general the primordial solar metallicity. Currently, the uncertainties in nuclear cross sections and neutrino oscillation parameters limit the precision of this approach; both uncertainties should be reduced to the level of 3% [55]. The $^3$He($\alpha,\gamma$)$^7$Be reaction has also important implications on the Big Bang nucleosynthesis. The LUNA I data (section 3) together with other input parameters have shown that the ratio $\eta$ of nucleons to photons in the early universe is given by $\eta_{BBN}$ = (5.75 ± 0.30)10$^{-10}$, [48, 56] in good agreement with $\eta_{WMAP}$ = (6.225 ± 0.170)×10$^{-10}$ derived from measurements of the cosmic background radiation using the Wilkenson Microwave Anisotropy Probe (WMAP) [57,58]. A detailed comparison of the abundances of the primordial elements (D, $^3$He, $^4$He, $^7$Li) from WMAP results and astronomical observations demonstrate a good agreement for the D and $^4$He abundance. However, the predicted abundance of $^7$Li is a factor 2 to 3 higher than observations. According to the Standard Model of Big-Bang nucleosynthesis, $^7$Li is produced by the $^3$He($\alpha,\gamma$)$^7$Be reaction followed by the electron capture of $^7$Be. Therefore this reaction rate is the necessary basis for possible solutions of this $^7$Li problem. The important energy range of the Big Bang nucleosynthesis for the $^3$He($\alpha,\gamma$)$^7$Be reaction is $E_0$ = 80 to 400 keV.

The absolute cross section of the $^3$He($\alpha,\gamma$)$^7$Be reaction was derived from the observed flux of capture $\gamma$-rays (prompt $\gamma$ method) as well as from the observed radioactivity of the residual nuclei $^7$Be, i.e. its electron capture to $^7$Li with $T_{1/2}$ = 53 days (activation method). There is an apparent discrepancy of about 15 % between the results obtained from both methods [59]. Its origin is presently not understood: it might be found either in the underestimation of systematic effects or in the presence of non electromagnetic transition (monopole) that can explain the 15 % larger value of the activation measurements with respect to the prompt $\gamma$-rays. The latter explanation seems to be unlikely according to calculations [60].

An improvement in the knowledge of the $^3$He($\alpha,\gamma$)$^7$Be reaction could come either from an independent direct detection of the $^7$Be reaction products conserving the kinematic information or new precision measurements using both activation and prompt $\gamma$ methods at the same time to reduce systematic errors. The second approach was followed at LUNA coupling high accuracy with low-energy data. The studies included an extended $^3$He recirculating gas target and a 137 % HPGe detector positioned at close distance to the interaction chamber. The intense $^4$He beam was focused on a copper beam stop that served as the end cap of a calorimeter as well as a catcher for the $^7$Be nuclei allowing for a later off-line measurement of the collected radioactivity. The prompt $\gamma$-rays of the $^3$He($\alpha,\gamma$)$^7$Be reaction (Q = 1.586 MeV) are located in the region of natural background. As mentioned in section 3 passive lead shielding is very effective in an underground laboratory (figure 4b), therefore a massive shield was built around both target chamber and detector. The shield consisted of several layers of lead bricks and an inner layer of Oxygen Free High Conductivity (OFHC) copper bricks (figure 12). Moreover all materials inside the shielding, i.e. target chamber and beam calorimeter, have been constructed from low background materials where possible. Figure 13 compares a background spectrum with that of prompt $\gamma$-rays obtained at the lowest energy of E = 93 keV. The achieved background reduction was 5 orders of magnitude in the $\gamma$-ray energy region below 2 MeV with respect to a spectrum taken without shielding in the underground laboratory [61]. The counting rate drops significantly towards the lowest beam energy and reaches a reaction cross section of about 200 pb. In order to maintain a good signal-to-noise ratio, the beam current should be as high as possible. However, in the case of intense beam currents (hundreds of μA) and gas targets, the beam heating effect has to be taken into account. Indeed, the ion beam heats the gas along its path reducing the effective gas density. This effect depends on the energy loss of the projectile inside the target matter, the beam intensity, and the target geometry. It had to be determined with high accuracy for a reliable absolute σ(E) measurement under these experimental conditions. For this purpose a dedicated monitor setup was placed inside the interaction chamber that allowed the observation of the elastically scattered alpha particles in a Si detector. A combination of a carbon foil and an aperture ensured that only double scattered particles from a well defined effective interaction region could reach the Si detector. In this way a reasonable scattering yield at these low energies was achieved. The beam heating effect was studied extensively over a wide energy range and at different pressures and beam currents, leading to a final precision of 2% in the determination of the target density [62].

A comparison between the results from the prompt γ-ray detection and activation was possible since both were carried out concurrently. This reduced significantly the systematic uncertainty between the two methods: the remaining uncertainties are the detection efficiency and the effective interaction length in the target. After each run the calorimeter cap was dismounted and placed in front of a heavily shielded 125% HPGe detector of the low-level laboratory of Gran Sasso [25, 63, 64]. This counting setup allowed to measure activities down to 25.3 ± 1.3 mBq, while a background sample – irradiated with $^4$He gas in the target chamber and a comparable irradiation time – resulted in a background value of lower than 0.1 mBq [64]. A sample spectrum with the γ-ray line of the $^7$Be electron capture at $E_\gamma$ = 478 keV is shown in figure 14.

The cross section was measured at E = 127 and 148 keV using activation technique only, while at E = 93, 106, and 170 keV the cross section was obtained using both techniques (activation and prompt γ-ray) (figure 15). The results from the two methods were consistent and did not show any discrepancy at the level of the achieved accuracy (4%) [65]. No discrepancy between the two techniques was found also by a recent experiment [67] where the $^3$He(α,γ)$^7$Be reaction was measured in the energy region between E = 0.35 and 1.2 MeV. These results excluded the contribution of non radiative contribution to the $^3$He(α,γ)$^7$Be cross section, at least at these energies. While the LUNA data covered the energy window of BBN, an extrapolation is still needed for the solar Gamow peak. An average S(0) value was obtained by combining the most recent data [66, 67] with the LUNA data [63, 64, 65]. The combined data were fitted rescaling two different theoretical curves (figure 15): a resonating-group calculation [68] and a direct capture model [69]. Following the approach indicated by [69], the final S(0) value was obtained from a weighted average of the extrapolated S(0) for each experiment and resulted in S(0) = 0.567 ± 0.018 ± 0.004 keV b where the last error value accounts for the uncertainty of the adopted theoretical model (figure 15) [70]. Note the systematic difference between the data sets (figure 15). Therefore, further improvements could come from new experiments exploring, with the same setup, the entire energy range from 0.1 to a few MeV, thus determining the S factor energy dependence. Additionally, new approaches are needed such as ERNA [71], a recoil mass separator at Bochum, allowing to detect directly the $^7$Be recoils; this new approach is subject to different dependencies compared to the existing data sets and may allow to reduce further the systematic error of the astrophysical S factor.

*4.3 The $^{14}$N(p,γ)$^{15}$O reaction*

The $^{14}$N(p,γ)$^{15}$O reaction (Q = 7.297 MeV) is the slowest reaction in the H-burning CNO-cycle, which determines thus the rate of this cycle. Its Gamow peak for massive stars, e.g. Asymptotic Giant Branch stars (AGB), is at energies higher than $E_0$ = 70 keV while the solar Gamow peak is at $E_0$ = 26 keV and requires an extrapolation of data obtained at higher energies. These extrapolations are hampered by the complex decay structure with different energy dependencies and the non-negligible contributions from several transitions, which sum up incoherently to the total cross section: the energy dependence of each single transition needs to be determined to high accuracy as well as the total cross section.

The cross section for energies between 100 and 400 keV is dominated by a resonance at $E_R$ = 259 keV, which decays through the excited states at $E_x$ = 5.18, 5.24, 6.17, 6.79 MeV and directly to the ground state of $^{15}$O. However, for energies lower than 100 keV a subthreshold resonance at $E_R = -507$ keV might contribute significantly to the total cross section. To evaluate the contribution of this subthreshold state it is of crucial importance to extend the experimental data to energies as low as possible. Thus, the LUNA experiment was divided in two phases. Firstly, high resolution HPGe detectors in close and far geometry to a solid TiN target were used to untangle the energy dependency of each transition; the γ-ray detection efficiency was of course low (about $10^{-4}$) and the experiment could not reach the lowest desirable energies. Secondly, a large volume 4π BGO summing detector in combination with a windowless gas target filled with natural nitrogen gas has been used, in order to extend the measurements as close as possible to the solar Gamow energy.

The experiment in both phases benefited from the advantages of the underground laboratory. In particular, in the study with the solid state target the γ-ray energies of the ground state transition as well as of the secondary cascade-transition are well above the natural γ-ray background, due to the high Q-value of $^{14}$N(p,γ)$^{15}$O and the high energy of the involved excited states in $^{15}$O. The natural γ-ray background above $E_\gamma = 4$ MeV is almost negligible (figure 16). But the measurements are limited by beam-induced γ-ray background, mainly from the reaction $^{11}$B(p,γ)$^{12}$C. In contrast, the primary transitions to excited states were located within the natural γ-ray background region, but the higher detection efficiency at low γ-ray energies partly compensates this problem. Thus, the γ-ray spectra allowed an analysis of both the primary and secondary transition for each cascade, leading not only to a high sensitivity but also to an internal efficiency check. The solid TiN target was produced by plasma deposition on a Ta backing and all γ-ray spectra were obtained at a detection angle of 55° minimizing angular distribution effects. The efficiency of the HPGe detector was determined for different distances between detector and target at the $E_R = 259$ keV resonance normalized with calibrated γ-ray sources, e.g. $^{137}$Cs and $^{60}$Co, placed at the target position. In the course of the experiment the parameters of the $E_R = 259$ keV resonance, i.e. strength, width, and branching ratios, were extracted with high accuracy (table 1). In the data analysis the correction of summing-in effects was treated with special care. Summing effects occur in a sizable way for a close geometry: the intensity of the weak ground state transition (1 - 2%) is heavily masked by summing contributions of cascade transitions leading to an incorrect ground state energy dependence of σ(E) at low energies in the earlier studies. All the observed ground state yields were corrected for summing-in, as well as the cascade transitions for summing-out. Data were taken in the energy window $120 < E < 370$ keV (figure 17) and, indeed, the measurement was limited by the increasing summing corrections becoming larger than the statistical error, although from the signal-to-noise ratio a lower energy seemed possible [72, 73].

Consequently, a new experiment was started to extend the data set to lower energies using a windowless $^{14}$N gas target coupled with a 4π BGO summing crystal. The setup was similar to that described in section 3.4. The 12 cm long target cell was placed inside the borehole (6 cm diameter) of the BGO crystal, ensuring ≈ 70% peak detection efficiency for 7 MeV γ rays. Typical gas pressure was 1 mbar. Prior to the actual experiment, the pressure profile along the target length was investigated with an identical chamber except for several apertures along the target length to allow for the pressure measurements at several distances. The beam-heating effect (section 4.2) reduced the target density by up to 15% compared to the static situation. The influence of this effect was studied with an indirect method exploiting the $E_R = 259$ keV resonance with a collimated 1×1" NaI detector mounted on a rail movable parallel to the beam axis [74]. Due to the high efficiency of the BGO detector and its 4π geometry, all the γ-ray transitions were mostly summed to a single summing peak around 7.5 MeV.

The measurement was carried out from $E = 70 – 233$ keV [74, 75]. The beam-induced background in the region of interest (ROI: 6.5-8 MeV) resulted from the Compton continuum of high-energy γ-rays from $^{15}$N(p,γ)$^{16}$O, $^{11}$B(p,γ)$^{12}$C, and $^{13}$C(p,γ)$^{14}$N. The number of events from $^{13}$C(p,γ)$^{14}$N ($E_\gamma \approx 7.7$ MeV) was evaluated by taking γ-ray spectra with helium instead of nitrogen as target gas. At the lowest beam energies, the main contribution to the background in the ROI was laboratory background due to (n,γ) reactions caused by neutrons from (α,n) reactions in the rocks (section 3.1). At $E = 70$ keV, with 49 days of running time, there were 11 counts/day from the reaction, 21 counts/day from laboratory background and 1 count/day from beam-induced background (figure 18). The background reduction due to the underground laboratory in the region of interest was about 3 orders of magnitude [76]. The low-energy limit is higher than the solar Gamow peak, but it is within the full Gamow peak for AGB stars (figure 19). The yield decrease from $E = 70$ keV (11 counts per day in the LUNA experiment) to 60 keV would decrease by an order of magnitude and another 5 orders of magnitude at 30 keV unreachable with the present techniques. Therefore, an extrapolation of the S(E) factor is still necessary.

The combined data sets of LUNA - together with previous data obtained at higher energies - showed that the influence of the subthreshold resonance was overestimated in the earlier extrapolations for the capture into the ground state, while capture into excited states was properly extrapolated. Thus, the true rate is reduced by about a factor of 2 for temperatures below $150 \times 10^6$ K [73]. It should be pointed out, that the reliable extrapolation of the LUNA II data to the solar Gamow peak required the combination of low-energy and high-energy data. This statement is generally valid for any extrapolation.

A comparison of the LUNA results with those obtained by [77] showed a good agreement in the overlapping energy region for all transitions, except for the ground state transition. Moreover, an R-matrix analysis of all available data for the ground state transition showed a large discrepancy due to the large uncertainties among the data sets in the region above the 259 keV resonance. As already pointed out the ground state data obtained in close geometry are largely influenced by summing-in effects. For this reason LUNA recently decided to perform an experiment in the energy range 300 to 400 keV using a BGO shielded Clover HPGe detector to reduce significantly the summing–in contributions. The results (figure 20) confirm the LUNA data and R-Matrix extrapolation [72, 73]. The previous discrepancy has been resolved, and the ground state capture no longer dominates the uncertainty of the total S-factor (table 1) [79].

**Table 1:** The excitation energies of $^{15}$O states and the branching ratios at the $E_R$ = 259 keV resonance; the last column gives the individual S factor extrapolated to zero energy.

| Transition | $E_X$ [keV] | branching ratios [%] | S(0) [keV b] |
|---|---|---|---|
| 5.18 | 5180.8 ± 0.2 | 17.1 ± 0.2 | 0.010 ± 0.003 |
| 5.24 | 5240.0 ± 0.3 | 0.6 ± 0.3 | 0.070 ± 0.003 |
| 6.17 | 6172.3 ± 0.2 | 57.8 ± 0.3 | 0.08 ± 0.03 |
| 6.79 | 6791.0 ± 0.2 | 22.9 ± 0.3 | 1.20 ± 0.05 |
| g.s. | 7556.4 ± 0.6 | 1.6 ± 0.1 | 0.20 ± 0.05 |

The rate reduction by a factor 2 had several astrophysical implications, which are described in the following discussion.

(i) In the sun the CNO-cycle contributes 0.8 % - instead of 1.6 % - to the total energy production [80, 81] and thus the corresponding flux of solar CNO neutrinos (i.e. below $E_\nu$ = 7 MeV) is reduced by this factor [82], which is of high relevance for the Borexino neutrino detector [83, 84]. The impact of nuclear cross sections on the physics of the sun has been discussed in section 4.2, e.g. the possibility to exploit solar neutrinos to determine the primordial solar core abundances of C and N, and more general the primordial solar metallicity.

(ii) Globular clusters represent the oldest resolved stellar populations. Their age coincides with the time elapsed since the epoch of formation of the first stars in the universe and provides an independent check of the reliability of standard (and non-standard) cosmological models. Among the various methods to determine the age of stellar clusters, the most reliable and widely adopted method is that based on the measurement of the luminosity of the turnoff point, i.e. the bluest point in the main sequence. This dating technique requires the knowledge of distance, light extinction along the line of sight, and chemical composition of the cluster. In addition, a reliable theoretical calibration of the turnoff luminosity-age relation is needed. This relies on our current knowledge of the physical processes of energy generation (e.g. nuclear reactions) and transport (e.g. opacity) taking place in H-burning low mass stars. The main sequence stars presently observed in globular clusters have masses smaller than that of the sun. These low-mass stars burn H in the centre, mainly through the pp-chain. However, towards the end of their life, when the central hydrogen mass fraction becomes smaller than about 0.1, the nuclear energy released by the H-burning becomes insufficient and the stellar core contracts to extract energy from its own gravitational field. Then, the central temperature and density increase and H-burning switches from the pp-chain to the more efficient CNO-cycle. Thus, the departure from the Main Sequence is powered by the CNO-cycle, whose bottleneck is the $^{14}$N(p,γ)$^{15}$O reaction. The luminosity of the turnoff point depends on the rate of this key reaction: the larger the rate, the fainter the turnoff. In contrast, the total lifetime is only marginally affected by this change, because the lifetime is mainly determined by the rate of the slowest pp-chain reaction, $p(p,e^+\nu_e)d$. As a consequence, a decrease of the CNO rate implies a brighter turnoff point for a given age, or a younger age for a given turnoff luminosity. Therefore, using the new lower reaction rate the age of globular clusters is reduced by about 1 billion years to 14 billion years, in better agreement with the age of the universe from other determinations [85]. In particular, the Globular Cluster age is compatible with the results of WMAP, a satellite telescope measuring the fluctuation of the microwave background radiation. WMAP has accurately measured the fundamental cosmological constants, deriving the age of the universe from the $\Lambda$ Cold Dark Matter model to $13.7 \pm 0.2$ Gy [57].

(iii) It is widely accepted that AGB stars are important for the chemical evolution of galaxies. However, there existed a long standing problem between observation and stellar models, i.e. an over-production of all chemical elements above C by a factor 2 or more [86]. This discrepancy has been resolved by the lower rate of $^{14}N(p,\gamma)^{15}O$. During the AGB phase the energy required to balance the surface irradiation is mainly provided by the H-burning shell, located just below the inner border of the convective envelope. This situation is recurrently interrupted by the growing up of thermonuclear runaways (thermal pulses) in the He-shell; as a consequence of a TP, the He-intershell becomes unstable to convection for a short period, the external layers expand and the H shell burning temporarily dies out. Then, the convective envelope can penetrate into the C-rich He-intershell (third dredge up episode), transporting to the surface the elements previously synthesized in the internal layers. The smaller the $^{14}N(p,\gamma)^{15}O$ rate, the smaller the helium production rate and the later the ignition of the He-shell flash. As a consequence the flash will be more violent and the forthcoming dredge-up is more efficient compared to a scenario with a larger $^{14}N(p,\gamma)^{15}O$ rate. Finally, a more efficient dredge-up leads to a larger envelope enrichment, in particular higher $^{12}C$ and metal abundances, which will pollute the interstellar medium [86].

*4.4 The $^{25}Mg(p,\gamma)^{26}Al$ reaction*

Observations from satellites [87, 88] have discovered a $\gamma$-ray line at 1809 keV, which arises from the $\beta$-decay of $^{26}Al$ to $^{26}Mg$ ($T_{1/2} = 7\times10^5$ yr). The intensity of the line corresponds to about 6 solar masses of $^{26}Al$ in our galaxy. Moreover, the presence of $^{26}Al$ in the interstellar medium has been determined from the observation of $^{26}Mg$ isotopic enrichment (extinct $^{26}Al$) in carbonaceous meteorites [89]. While the observations from COMPTEL and INTEGRAL provided evidence that $^{26}Al$ nucleosynthesis is still active on a large scale, the Mg isotopic variations show that $^{26}Mg$ must have been produced within the last 4.6 billion years (time of the condensation of solar-system material). Any astrophysical scenario for $^{26}Al$ nucleosynthesis must be concordant with both observations.

The nuclides $^{26}Al$ are produced mainly via the $^{25}Mg(p,\gamma)^{26}Al$ capture reaction. The most important site for the activation of this reaction is the hydrogen-burning shell (HBS), which may be active in off main sequence stars of any mass. In particular, the Mg-Al cycle is at work in the hottest region of the HBS, close to the point of the maximum nuclear energy release. In addition, $^{25}Mg(p,\gamma)^{26}Al$ may be also active within the carbon-burning regions in massive stars. In the HBS, the $^{25}Mg(p,\gamma)^{26}Al$ reaction starts when the temperature exceeds about $30\times10^6$ K and between $40 < T/10^6$ K $< 60$ (corresponding to a Gamow energy of about $E_0 = 100$ keV) almost all the original $^{25}Mg$ is converted into $^{26}Al$. At higher temperatures, the destruction of $^{26}Al$ by $^{26}Al(p,\gamma)^{27}Si$ and the refurbishment of $^{25}Mg$ by the sequence $^{24}Mg(p,\gamma)^{25}Al(\beta^+)^{25}Mg$ begins to play a relevant role. Once the HBS advances in mass, the $^{26}Al$ is accumulated within the H-depleted core. A convective dredge up coupled to a huge stellar wind or an explosive ejection of the freshly synthesized material are needed in order to make the $^{26}Al$ ashes an astronomical "observable". However, these processes must occur before the extinction of $^{26}Al$ through its radioactive decay.

The existence of an active HBS is a common feature in stellar evolution and different classes of stars can have these characteristics: low mass AGB, massive AGB, Novae, Core Collapse Supernovae and Wolf-Rayet stars. Stellar nucleosynthesis studies predict that 30 to 50% of $^{26}Al$ is produced in the HBS of massive stars (core collapse supernovae or WR-stars). The source of the remaining contribution is unknown and a more precise knowledge of the relevant reaction rates certainly will help in reducing the range of free parameters. Such a project was recently started by the LUNA collaboration.

The reaction $^{25}$Mg(p,γ)$^{26}$Al (Q = 6.306 MeV) is dominated by narrow resonances which have been found down to $E_L$ = 190 keV. From the known level structure of $^{26}$Al one expects low-lying resonances at E = 93, 109, and 130 keV, among which the 93 keV resonance appeared most important. Indeed, the 93 and 130 keV resonances were found at LUNA II (figure 21) using the 4π BGO crystal and a $^{25}$Mg solid target [90]. The data allow for improved astrophysical calculations of the origin of the 1809 keV line as well as of the chemical evolution of galaxies.

Moreover, the reaction $^{25}$Mg(p,γ)$^{26}$Al is a good example to demonstrate the influence of beam-induced background. For the study of a given reaction with particles of charge $Z_1$ and $Z_2$ in the entrance channel, the high sensitivity of the cross section to the height of the Coulomb barrier produces a beam-induced background, which can be much larger than the signal of interest. This is the case particularly for contaminations in the ion beam or in the target if their nuclear charges are smaller than $Z_1$ and/or $Z_2$. Due to the high energy resolution of HPGe detectors in these studies, some contributions from beam-induced background can be tolerated [61], while measurements with high-efficiency scintillator detectors, e.g. the 4π BGO detector, are limited by such kind of background in an underground laboratory. In the absence of cosmic-ray background even tiny contaminations of light elements are observable in a γ-ray spectrum obtained at ultra-low energies. Furthermore, the proton capture on these elements often have high Q-values as in the case of the reaction $^{7}$Li(p,γ)$^{8}$Be (Q = 17.255 MeV), $^{11}$B(p,γ)$^{12}$C (Q = 15.957 MeV), and $^{18}$O(p,γ)$^{19}$F (Q = 7.995 MeV). The capture reactions on $^{11}$B and $^{18}$O have strong resonances at $E_R$ = 149 and 143.5 keV, respectively, which almost exclude this energy region for precise cross section measurements. The reaction $^{7}$Li(p,γ)$^{8}$Be proceeds through a direct capture process with a fairly high cross section and, indeed, the γ-ray signal of this reaction can be observed in the spectrum of the $^{25}$Mg(p,γ)$^{26}$Al measurement at $E_p$ = 100 keV above the region of interest (figure 21). Another important contaminant reaction is $^{19}$F(p,αγ)$^{16}$O with a single γ-ray line at $E_γ$ = 6.13 MeV, often dominating the beam-induced background in the energy region above $E_p$ = 300 keV; it is almost absent at lower energies since the lowest resonance of this reaction is located at $E_R$ = 224 keV. These traces of light elements are impurities in the target or in the backing materials, often unavoidable in the production process of metallic materials. Nevertheless, it is the responsibility of the experimentalists to avoid any unnecessary contaminations of any material in the target chamber. We discuss another example of target contamination in section 5.2.

The precision and reliability of absolute values for the cross section or resonance strength require unusual efforts in the determination of all quantities entering the determination of these values. To discover systematic errors one has to use different approaches. Recently, the AMS approach (AMS for Accelerator Mass Spectrometry) has been used as such an alternative approach [91]. The $^{26}$Al nuclei represents in principal an ideal case for such a study since $^{26}$Al is radioactive with a reasonably long half life and the corresponding stable isobar $^{26}$Mg does not form negative ions. The chemistry and AMS procedure are now standard techniques and well established. However, the quality and stability of the target remains a challenge in these studies and reactions dominated by narrow resonances, such as $^{25}$Mg(p,γ)$^{26}$Al, are especially difficult. The observed yield, in γ-ray spectroscopy as well as AMS, for a particular narrow resonance (thick-target yield) depends linearly on the inverse of the effective stopping power. This important quantity includes the active nuclei, i.e. $^{25}$Mg, but also inactive nuclei in the target, for example isotopic impurities or other contaminants such as oxygen. Moreover, it is well known in experimental nuclear astrophysics that a solid state target under heavy proton bombardment changes its stoichiometry in the course of the measurement and a frequent control of the target quality is absolutely necessary. A standard technique, the normalization to a precise known and stronger resonance at higher energy of the same reaction, cannot be applied in an AMS experiment since this normalization would lead to the production of a much larger number of reaction products, sometimes several orders of magnitude: such a sample cannot be used for AMS. Nevertheless, the underground laboratory might often offer a solution. It is almost impossible to produce isotopically pure targets. For most of the elements these isotopic impurities will be in the order of several percent and a normalization to a known resonance of a reaction on one of the isotopes might be feasible if the quantity of the impurity is known and the detection setup is sensitive enough. In the case of the $^{25}$Mg(p,γ)$^{26}$Al reaction an enriched $^{25}$Mg target contained about 1.5% $^{24}$Mg and a low energy $^{25}$Mg(p,γ)$^{26}$Al AMS measurement can be normalized to the $E_R$ = 224 keV resonance in $^{24}$Mg(p,γ)$^{25}$Al.

Finally, it should be noted that isotopic anomalies - such as $^{26}$Mg/$^{24}$Mg - have been found in meteoritic samples for many elements across the periodic table. They give clear evidence that the mixing of the stellar ashes in the interstellar space has not been complete and thus the anomalies tell us something about the "last salting" of the stellar debris. Among the anomalies of the lighter nuclides such as Ne-E (nearly pure $^{22}$Ne, E for extraordinary), there are open questions with regard to their nuclear astrophysics origin such as the NeNa and MgAl cycles. This cycles may not exist at all as cycles [3]. Thus, further work in an underground laboratory is needed. The LUNA collaboration submitted last year a 5 year program to measure some of these reactions with the present 400 kV accelerator. The Italian Institute for Nuclear Physics (INFN) approved the proposal and the new program has already started with the measurement of $^{15}$N(p,γ)$^{16}$O. The other reactions on the list are: D(α,γ)$^6$Li, $^{17,18}$O(p,γ)$^{18,19}$F, $^{22}$Ne(p,γ)$^{23}$Na, and $^{23}$Na(p,γ)$^{24}$Mg.

**5. The future of underground laboratories**

Although LUNA is yet unique in the world and will continue for several more years, it is hoped that similar facilities will be created in time worldwide. Each fusion reaction takes for its complete study a couple of years. Since there are many reactions to be reinvestigated, there is a clear need for additional underground facilities. We discuss in the following subchapters three illustrating examples for a renewed and improved study of important nuclear reactions induced by charged particles.

*5.1. The holy grail $^{12}$C(α,γ)$^{16}$O of Willy Fowler*

The unknown rate of the $^{12}$C(α,γ)$^{16}$O reaction is responsible for one of the most important uncertainties in nuclear astrophysics today. The $^{12}$C(α,γ)$^{16}$O process influences the evolution of stars, i.e. all stars with M > 0.55M$_\odot$ burn He in its core, essentially in two basic manners. First, the He-burning is directly affected since the reaction operates primarily in this evolutionary phase, and, second, the subsequent stellar evolution is governed by the chemical composition of the matter left by the He-burning. The primary reaction of the He-burning, the triple-α process, starts after the stellar core temperature exceeds $10^8$ K. The synthesized $^{12}$C can capture another α particle forming an oxygen nucleus and by another α-capture the reaction $^{16}$O(α,γ)$^{20}$Ne can occur, which is however for several reasons negligible. Therefore, the blocking of further α captures for standard He-burning conditions implies that $^{12}$C(α,γ)$^{16}$O and the triple-α process are the most important reactions. The Q values of both processes are very similar, but the $^{12}$C(α,γ)$^{16}$O reaction consumes only one α particle and, thus, determines the ashes of this phase: the larger the $^{12}$C(α,γ)$^{16}$O reaction rate the longer the He-burning time and, consequently, the amount of $^{16}$O.
The final C abundance at the end of helium-burning has important consequences for the following astrophysical scenarios. (i) Composition of type II supernovae (SN) ejecta [92, 93]; the intermediate-light elements Ne, Na, Mg and Al scale directly, while the elements produced by the explosive burning phases, i.e. complete and incomplete explosive Si-burning, explosive O-burning, and explosive Ne-burning, scale inversely with the C abundance left by the He-burning. Moreover, the structure of the progenitor star is more compact and the mass of the explosive core lower for a larger $^{12}$C(α,γ)$^{16}$O rate. Thus, there is more energy available for the shock wave induced by the core collapse [11]. (ii) The rise time of the type I SN light curves; the maximum luminosity and the kinetic energy of the SN are driven by the $^{56}$Ni abundance which is again proportional to the C abundance. Therefore, the lower the $^{12}$C(α,γ)$^{16}$O rate, the faster and brighter the light curve [94]. (iii) The white dwarf cooling sequences; CO white dwarfs undergo stable pulsations during their long cooling time. These pulsations, observed as variations in brightness, in principal provide constraints on the internal structure of these condensed objects, e.g. the chemical profile. In particular, an oxygen abundance was found, which is – according to present models – significantly larger than expected from the $^{12}$C(α,γ)$^{16}$O reaction rate [95]. However, large uncertainties in the efficiency of convection induced mixing make predictions of the central oxygen mass fraction rather uncertain. An accurate knowledge of the reaction rate could improve our understanding of the convection processes [96].
The $^{12}$C(α,γ)$^{16}$O capture process proceeds predominantly to the ground state of $^{16}$O with electric dipole (E1) and electric quadrupole (E2) amplitudes. The E2 part is of equal importance to the E1 part - a rare case -, since in this reaction with equal charge to mass ratios for the particles in the entrance channel there is no moving charge, and thus the E1 multipole is forbidden, i.e. it is at least strongly suppressed. Both amplitudes are influenced at low energies by subthreshold resonances at $E_x$ = 6.92 and 7.12

MeV. These are excited states lying somewhat below the threshold energy of the reaction, Q = 7.162 MeV; due to their finite widths caused by γ-ray decay, these states have a high-energy tail extending eventually above the threshold energy. Due to these subthreshold resonances and other resonances including their mutual interference effects, the energy dependences of the E1 and E2 cross sections are quite different. Furthermore, there could be capture processes to excited states of $^{16}$O, which again have different energy dependences compared to the capture into the ground state.

In order to extract the different multipole contributions, a measurement of the γ-ray angular distributions at each energy is necessary. It must involve several angles ranging from 0° to nearly 180°. With such data of sufficient statistics, the ratio of the cross sections $\sigma_{E1}(E)$ and $\sigma_{E2}(E)$ as well as their relative phase difference can be obtained from a fit of the angular distributions. Constraints on the phase shift in overlapping energy regions are given by elastic scattering data [97]. However, such capture experiments are extremely difficult below E = 1 MeV due to the low cross section as well as increasing problems with target purity and stability [98].

The present status of the data is shown in figure 22. The ERNA data for the total capture cross section [12] of $^{12}$C(α,γ)$^{16}$O - obtained from the $^{16}$O recoils - are deconvoluted into the individual capture amplitudes discussed above based on an R matrix calculation [99]. The individual contributions, i.e. the E1 and E2 amplitudes, are compared with direct γ-ray measurements [98], where the data set reaches a lower-limit of $E_L$ = 0.9 MeV, with large uncertainties and far away from the associated Gamow energy at $E_0$ = 0.3 MeV (for details, see [100]). Furthermore, the question of the $E_x$ = 6.05 MeV excited state as a possible third subthreshold resonance needs to be answered; this $J^\pi = 0^+$ state has a sizable alpha-cluster-structure much higher than for the $0^+$ ground state. This amplitude was observed with a fairly high cross section at energies above E = 2.2 MeV [101] and could amount to about 15% of the total capture cross section at $E_0$ according to an R matrix study of [101]. Finally, there is the open question of a monopole capture process (E0) into the ground state, which involves an s-partial wave in the entrance channel, while the E1 and E2 amplitudes involve p- and d-partial waves, respectively, with their respective centrifugal barriers. Although a recent theoretical work indicated that the latter process is unlikely [102], a direct E0 observation would require the development of an electron-positron spectrometer in close geometry.

In all direct studies the target preparation represents an important issue. The Stuttgart-Orsay group [98] was able to produce ultra-clean $^{12}$C targets with a $^{13}$C depletion – the most important impurity due to the $^{13}$C(α,n)$^{16}$O reaction – of 5 orders of magnitude. These targets with a thickness between $1\times10^{18}$ and $10\times10^{18}$ atoms/cm$^2$ could withstand $^4$He beam currents of up to 350 μA although strong target deterioration was observed and the targets were replaced after a loss of 20%. However, there is no advantage in increasing the target thickness beyond about $10^{19}$ atoms/cm$^2$ because the drop in cross section with decreasing energy has the consequence that only the surface layer contributes significantly to the reaction yield. Thus, in a renewed study of the $^{12}$C(α,γ)$^{16}$O reaction in direct kinematics – with a $^4$He beam – an improvement seems to be possible only with a larger, highly segmented detector array and larger solid angle coverage than in previous experiments. On the other hand an important step forward would be a γ-ray spectroscopy experiment in inverse kinematics, e.g. with an intense $^{12}$C beam on a windowless $^4$He jet gas target. Such an experiment – as shown in a previous study [103] – is mainly limited by statistics (beam intensity) and the cosmic-ray background. The beam-induced background would be strongly reduced due to the $^{12}$C beam and the detection of the high-energy γ-rays would take again full advantage of the underground environment. Moreover, an experiment in inverse kinematics will lead to different systematic dependencies and, therefore, to additional and independent information.

The appropriate accelerator for this kind of experiment would be either a small 2 MV tandem accelerator with a standard sputter ion source, although no larger improvement compared to the state of the art techniques can be expected from this approach, or a 3 MV single-stage accelerator with an ECR ion source producing intense beams with high charge states. This accelerator type has been recently commissioned at the Jannus Facility in Saclay, France, called "Éphiméthée" [104] and should deliver a $^{12}$C beam with a current larger than 300 particle μA, thus, comparable with the intensity of available $^4$He beams in present studies. Such currents have in turn significant effects on the target. In the case of a windowless gas target – as mentioned already above – the ion beam heats the gas along its path reducing the effective gas pressure. This effect has been observed and must be taken into account, but it is less pronounced if not absent for a jet gas target [105] because of the frequent change of the target gas in the jet. Indeed, this is a clear advantage of this approach compared to a solid target because not only the stability of the target is important but also the purity. In order to reduce the background in case of a solid target the backing material was covered with a gold layer and the $^{12}$C target atoms were implanted into this layer [98]. Therefore, the energy loss in the target is mainly coming from the gold nuclei while in the case of the pure $^{4}$He gas the energy loss arises only from these target nuclei resulting in a higher reaction yield for the same target density.

Both approaches, direct and inverse studies, have their difficulties. Independent of the choice of the experiment their solution will be a clear challenge for the experimentalists; but they will benefit strongly by a placement in an underground laboratory. The Q-value of the reaction is rather high (comparable to the $^{14}$N(p,γ)$^{15}$O reaction); thus, the γ-ray energy of the ground state transition is well above the environmental background and could be observed in a nearly γ-ray background free environment (section 4.3). A measurement of the γ-ray angular distribution with a lower energy limit closer to the Gamow energy than presently available, i.e. E ≈ 0.5 MeV, seems to be possible provided the beam-induced background can be reduced to a negligible level.

*5.2. Carbon-burning*

The reactions $^{12}$C($^{12}$C,α)$^{20}$Ne (Q = 4.617 MeV) and $^{12}$C($^{12}$C,p)$^{23}$Na (Q = 2.241 MeV) are referred to as carbon-burning in stars [3, 106]. Carbon-burning represents the third stage of stellar evolution for massive stars (M > 8M$_{sun}$) and proceeds mainly through the $^{12}$C + $^{12}$C fusion reactions and, to a lesser extent, the $^{12}$C + $^{16}$O processes. The above reactions represent key reactions since they influence not only the nucleosynthesis of $^{20}$Ne and $^{23}$Na but also the explosion of the star, i.e. type Ia supernovae. These supernova explosions are driven by carbon ignition in cores of accreting massive C/O white dwarfs [107]. The burning process proceeds from the carbon ignition region near the centre of a white dwarf by detonation or deflagration through the entire white dwarf body. The ignition conditions and time scales are defined by the $^{12}$C + $^{12}$C reaction rate [108]. Thus, the cross sections of these reactions must be known with high accuracy down to the Gamow energy $E_0$ = 1.5 MeV (for a temperature of 5 × 10$^8$ K corresponding of the stellar core carbon-burning phase temperature). Using charged-particle or γ-ray spectroscopy the reactions were investigated over a wide range of energies down to E = 2.1 MeV [109, 110 and references therein]. However, below E = 3.0 MeV the reported cross sections are rather uncertain, because at these energies the presence of $^{1}$H and $^{2}$H contamination in the carbon targets often hampered the measurement of the $^{12}$C + $^{12}$C process in both particle and γ-ray studies. For example, in γ-ray spectroscopy the transitions from the first excited state in $^{20}$Ne (E$_γ$ = 1634 keV) and $^{23}$Na (E$_γ$ = 440 keV) were normally the prominent lines in the γ-spectra but at low energies their observation suffered from an intense background from the E$_γ$ ≈ 2.36 MeV line of $^{1}$H($^{12}$C,γ)$^{13}$N and the E$_γ$ = 3.09 MeV line of $^{2}$H($^{12}$C,pγ)$^{13}$C. Thus, improved studies required C targets with an ultra-low hydrogen contamination.

Recently, such an improved experiment with a HPGe detector in close geometry and surrounded by a 15 cm thick lead shield has been performed at the 4 MV Dynamitron tandem at the Ruhr-Universität Bochum [109]. The intense $^{12}$C beam (30 particle µA) heated the C target to a temperature of 700 °C leading to a rapid decrease of the hydrogen contamination in the target to a negligible level: the γ-spectra were now dominated by the lines of interest at 440 and 1634 keV. The present results are illustrated in figure 23 in the form of the modified astrophysical S(E)$^*$ factor defined by σ(E) = S(E)·E$^{−1}$ exp(−87.21E$^{−1/2}$ − 0.46E) with E in units of MeV. The resonance structure continues down to the low-energy limit, E$_L$ = 2.10 MeV, where a strong resonance was found at E$_R$ = 2.14 MeV [109]. In figure 24 the total S$^*$factor is displayed, which is mainly the sum of the α and p channels while the neutron channel is negligible at energies below E = 5 MeV [111]. The data are compared with different phenomenological calculations mainly based on various nuclear potentials [111, 112, 113, 114]. A detailed discussion of the theoretical models can be found in [110, 114]. However, most of the models seem to overestimate the non-resonant S$^*$ factor at low energies and clearly new measurements at energies below 2 MeV will improve the situation. It should be noted that no theory exists yet for the prediction of the location and strength of the $^{12}$C+$^{12}$C resonances. Moreover, since there are resonances all over the measured energy range, it is quite likely that a resonance exists near the centre of the Gamow peak which could completely dominate the reaction rate for astrophysical scenarios.

The γ-ray spectroscopy has the underline{disadvantage} that the transitions $\alpha_0$ and $p_0$ directly into the ground state of $^{20}$Ne and $^{23}$Na, respectively, cannot be observed. Direct particle identification would allow for measuring the total cross section for each channel. Such a measurement could be performed with standard silicon surface barrier detectors in close geometry to the target and under backward angles with respect to the beam direction. Furthermore, a compact $\Delta$E-E ionization chamber would allow for particle identification, e.g. $\alpha$ and p. In this geometry background problems arise from elastically scattered $^{12}$C nuclei. In addition, a close distance to the target leads to a heating of the detectors due to the heat radiation of the target under high intensity $^{12}$C beam. This effect influences drastically the performance of the particle detector. On the other hand an ion beam intensity of about 300 particle μA $^{12}$C – a factor 20 to 50 more than in previous experiments, e.g. [115] – would allow to mount the particle detectors in a far geometry without a significant loss of count rate compared to previous experiments but avoiding the problems of detector heating. However, such a particle spectroscopy measurement would only benefit from an underground laboratory if the two detectors of the $\Delta$E-E telescope are very close to each other and the detectors have a large surface area as in the case of the $^3$He($^3$He,2p)$^4$He LUNA experiment (section 3.2). An important issue in the particle spectroscopy is the background discrimination in particular in case of a low reaction yield: a distinguished background detection for heavy ion reactions is difficult if not impossible. In turn, this is an advantage of the γ-ray spectroscopy which provides a clear and unique signal if the signal-to-noise ratio allows for a precise observation of the signal. Therefore, the C+C fusion reactions are now an excellent case for experimental studies using a future underground facility, such as a 3 MV high-current, single-stage accelerator with an ECR ion source as discussed above.

The possible improvement for a γ-ray spectroscopy measurement of the $^{12}$C+$^{12}$C fusion reactions in an underground laboratory can be estimated from the LUNA $^3$He($\alpha$,γ)$^7$Be experiment presented in section 4.2. The detection setup of this experiment, in particular the arrangement of the passive shielding, was adapted to meet the special requirements for a measurement of γ-ray energies below 2.5 MeV. The environmental γ-ray background in this γ-ray energy region is even in an underground laboratory relatively high (figure 4a) compared to the reduction of cosmic-ray background. An advantage could come from the potential of salt mines where typically the unshielded γ-ray background provided by the salt environment is significantly reduced with respect to sedimentary rocks, including the $^{40}$K γ-ray line. This reduction approximately amounts to one order of magnitude, e.g. between the Boulby underground laboratory, U.K. (salt and potash mine), and Gran Sasso (sedimentary rock). The reduction could be larger for sites located in salt layers with a very low $^{40}$K content. Unfortunately, most of such sites are at depths less than 300 m and thus with an insufficient shielding against cosmic-ray background. However, in a well-shielded detector setup the actual environmental background is only one of the components determining the final background in the detector. Other important components are cosmic-ray induced reactions in the shielding material, i.e. in lead and copper, purity of the shielding material, and in particular beam-induced background (figure 21). Nevertheless, the achievements in terms of background reduction of the LUNA $^3$He($\alpha$,γ)$^7$Be prompt γ-ray experiment (figure 4b) can be considered as a standard for future nuclear cross section measurements. A similar or improved γ-ray background is probably within reach for all future underground accelerator laboratories at depths comparable to Gran Sasso.

Hence, the natural γ-ray background near the relevant 1634 keV γ-ray line of the $^{12}$C($^{12}$C,$\alpha$)$^{20}$Ne reaction can be used to calculate the sensitivity limit. The background for the HPGe detector in the well-shielded setup of the LUNA experiment is reduced by a factor 800 compared to the shielded setup in the recent Bochum experiment [109]. The $^{12}$C+$^{12}$C experiment could be performed in a similar way as previously with a single detector in close geometry. In a recent work [149] a hypothetical resonance at E = 1.5 MeV has been proposed based on the carbon ignition temperature in superburst models. The proposed experiment with a high intensity $^{12}$C ion beam, reasonable experimental parameters, and the above background assumption would have a resonance strength detection limit of at least 1 μeV for a resonance at E = 1.5 MeV. Thus, the hypothetical resonance of [149] would be within reach in such a measurement.

Unfortunately, a proper passive shielding similar to this experiment for an improved detection system such as a Ge crystal ball would lead to handling problems and large experimental difficulties. This approach could take advantage of the potential of salt mines due to the lower unshielded γ-ray background and the shielding requirements may be reduced. In summary, in both approaches combined with an ion beam current of 300 particle µA and long running times it appears possible to perform measurements over the energy range of the Gamow peak in an underground laboratory.

*5.3. Neutron sources for the trans-iron nucleosynthesis*

More than 50 years ago the synthesis of the trans-iron elements by neutron capture reactions on Fe seed nuclides was proposed by Burbidge, Burbidge, Fowler, and Hoyle [1] and independently by Cameron [116]. In spite of many theoretical efforts since then an exhausting explanation of how the neutron-capture nucleosynthesis operates in stars of different mass and composition is still missing. In a neutron rich environment the nucleosynthesis path follows – starting from the seed nuclides – towards the neutron drip line and eventually to higher elements in a series of neutron-captures and β decays. After the neutron flux vanishes all radioactive isotopes along the path decay until they reach a stable nucleus in the valley of stability. The solar abundance pattern for trans-iron elements cannot be explained by a single process: the nucleosynthesis proceeds by the so-called s-process – s stands for a slow neutron capture rate compared to the interim radioactive half life – and the r-process – rapid neutron capture rate compared to the half life. In the s-process, all isotopes with mass numbers between 60 and 90 are created by the so-called weak component, whereas heavier isotopes (from Kr to Bi) originate in the main and strong components. The weak component is active during the hydrostatic evolution of massive stars (core and shell He- and C-burning). Instead the production site of the main and the strong components are low-mass thermally pulsing AGB stars. For further details see the review articles [117, 118] and references therein. While the general physics of the s-process is relatively well understood, the astrophysical sites of the r-process are currently still under debate, even if there are strong evidences for a connection to core collapse supernovae [118] (figure 25).

The major source of the necessary neutrons is believed to arise from the $^{13}$C(α,n)$^{16}$O and $^{22}$Ne(α,n)$^{25}$Mg reactions in a He-burning environment, with a Gamow energy near 300 keV [1-3, 116]. In particular in low-mass AGB stars the s-process nucleosynthesis occurs in a thin radiative layer at the top of the He-intershell during the relatively long interpulse period – the time interval between two subsequent thermal pulses –, when the temperature ranges between $80\times10^6$ and $100\times10^6$ K. In such an environment the neutron density is low, $10^6$ to $10^7$ cm$^{-3}$, and is provided by the $^{13}$C(α,n)$^{16}$O reaction. A second neutron burst is released by a marginal activation of $^{22}$Ne with alpha particles in the convective thermal pulse of an AGB star. This short neutron exposure with a high peak neutron density, $10^{11}$ cm$^{-3}$, modifies the final s-process composition close to critical branching points located along the s-process path [118]. Moreover, the $^{22}$Ne(α,n)$^{25}$Mg reaction is the most important neutron source for the s-process in massive stars, i.e. M > 10 M$_\odot$, the production site of the solar weak component.

Both (α,n) reactions have been studied over a wide energy range, but none of the existing experimental investigations has reached the relevant astrophysical energies. Thus, for the $^{13}$C(α,n)$^{16}$O (Q = 2.216 MeV) reaction, the present low energy limit in a direct experiment is $E_L$ = 270 keV [120], while the Gamow peak is below 200 keV. The influence of a subthreshold state in $^{17}$O at $E_R$ = -0.3 MeV may suggest an increase of the S factor towards lower energy (figure 26) [120]. However, the contribution of this subthreshold resonance is presently heavily under debate. In recent indirect measurements [121, 122, 123] using various transfer reactions the Asymptotic Normalization Coefficient (ANC) for this state was determined. The extrapolation of the astrophysical S factor to the relevant energy based on these α-particle ANC results differ by a factor of 3 to 5 (figure 26), but always smaller than the reaction rate given in the NACRE compilation [124]. The observed differences between the results of these experiments show the difficulties associated with indirect methods. The differences should be thoroughly studied and understood.

The reaction rate of the $^{22}$Ne(α,n)$^{25}$Mg reaction (Q = -0.478 MeV) at stellar temperatures is characterized predominantly by a resonance at E = 0.70 MeV. However, a contribution by another α-unbound natural parity state at $E_x$ = 11.15 MeV in $^{26}$Mg is likely [125, 126]. In a recent work [127] only upper limits could be extracted below the resonance at E = 0.70 MeV and therefore the reaction rate at lower energies remains uncertain (figure 27).

Both reactions represent excellent cases for renewed studies in an underground laboratory. Neutron detectors with a nearly 50% detection efficiency have been developed in the past years [127], but they are sensitive to neutrons created by cosmic-rays in the environment requiring thus an underground surrounding including a shielding, such as a 1 m thick water housing around the whole setup to prevent neutrons from (α,n) reactions in the walls induced by natural α-radioactivity (section 3.1).

A preliminary background study in the Boulby underground laboratory with the neutron detector of [127] and without any further passive shielding has been performed recently [137]. This study suggests a background reduction of at least an order of magnitude for a measurement in the salt environment of the underground laboratory compared to earth's surface. The $^3$He counters used for the neutron detection of this detector are not made from low-level materials. This feature did not hamper the previous measurements on earth's surface. However, at the neutron background level of an underground laboratory the intrinsic background of the $^3$He counters would presently limit a measurement with this neutron detector. A new design of a neutron detector should certainly take this into consideration. The knowledge for the construction of low-level $^3$He counters is available from the SNO detector [138]. A total neutron background reduction of more than 2 orders of magnitude in a typical detector located in an underground laboratory compared to earth's surface appears feasible.

The possible improvement for a measurement of the $^{22}$Ne($\alpha$,n)$^{25}$Mg reaction in an underground laboratory – assuming a conservative background reduction of a factor 50 – is demonstrate in figure 27 by the horizontal dashed line. This line represents the sensitivity limit for a 3$\sigma$ upper limit under the mentioned neutron background in the detector. Reasonable assumptions for the experimental parameters, i.e. target density and beam intensity, and a measurement time of 5 days per data point were used for the calculation. The detection limit for the undiscovered resonance at E = 0.537 MeV would be reduced significantly.

**6. Epilogue**

The work of LUNA demonstrated the research potential of an underground laboratory for the field of nuclear astrophysics. Several key reactions could be studied at LUNA, some directly at the Gamow peak for the solar hydrogen burning.

There exist proposals for new underground laboratories such as in England, Spain, Romania, and USA [148, 128, 129]. To reduce significantly the cosmic-ray background events at higher $\gamma$-ray energies in the detectors, the laboratories should be several hundred meters underground. An underground laboratory has its special virtue for gamma- and neutron-spectroscopic studies. The natural background in the environment is minimized naturally in a salt mine but one can also install - in a rocky environment - a special salt or water house to absorb the radiation from the rocks. There should be enough space available to install sophisticated detection systems such as a crystal ball. Other rooms with underground observations such as neutrino detectors should be at a sufficient distance such that possible neutrons created in the accelerator laboratory represent no restriction. The accelerator may include also a pulsed beam technique for a further reduction of room or cosmic-ray background. Finally, there should be an easy access to the laboratory.

Impressive progress has been achieved in the knowledge of nuclear reaction rates. This includes improved R-matrix analyses for extrapolation of data to stellar energies. However, there remains much critical work to be done in the future to arrive at reliable data for many key reactions and processes. New techniques and approaches such as LUNA continue to be developed for this purpose and research in nuclear astrophysics will remain an exciting pursuit for many years to come.

**Acknowledgment**

This work has been done in the framework of the international LUNA collaboration. In spite of the long list of people who contributed to the studies in different phases over the years the authors are grateful in particular to: D. Bemmerer, C. Casella, F. Confortola, Gy. Gyürky, A. Lemut, and B. Limata. In addition, the authors would like to thank H.-P. Trautvetter and O. Straniero for fruitful discussions during the preparation of the manuscript. The present work has been supported by INFN and in part by the German Federal Ministry of Education and Research (BMBF, 05CL1PC1-1), the Deutsche Forschungsgemeinschaft (DFG, Ro 429/41), the Hungarian Scientific Research Fund (T42733 and T49245), and the EU (ILIAS-TA RII3-CT-2004-506222).

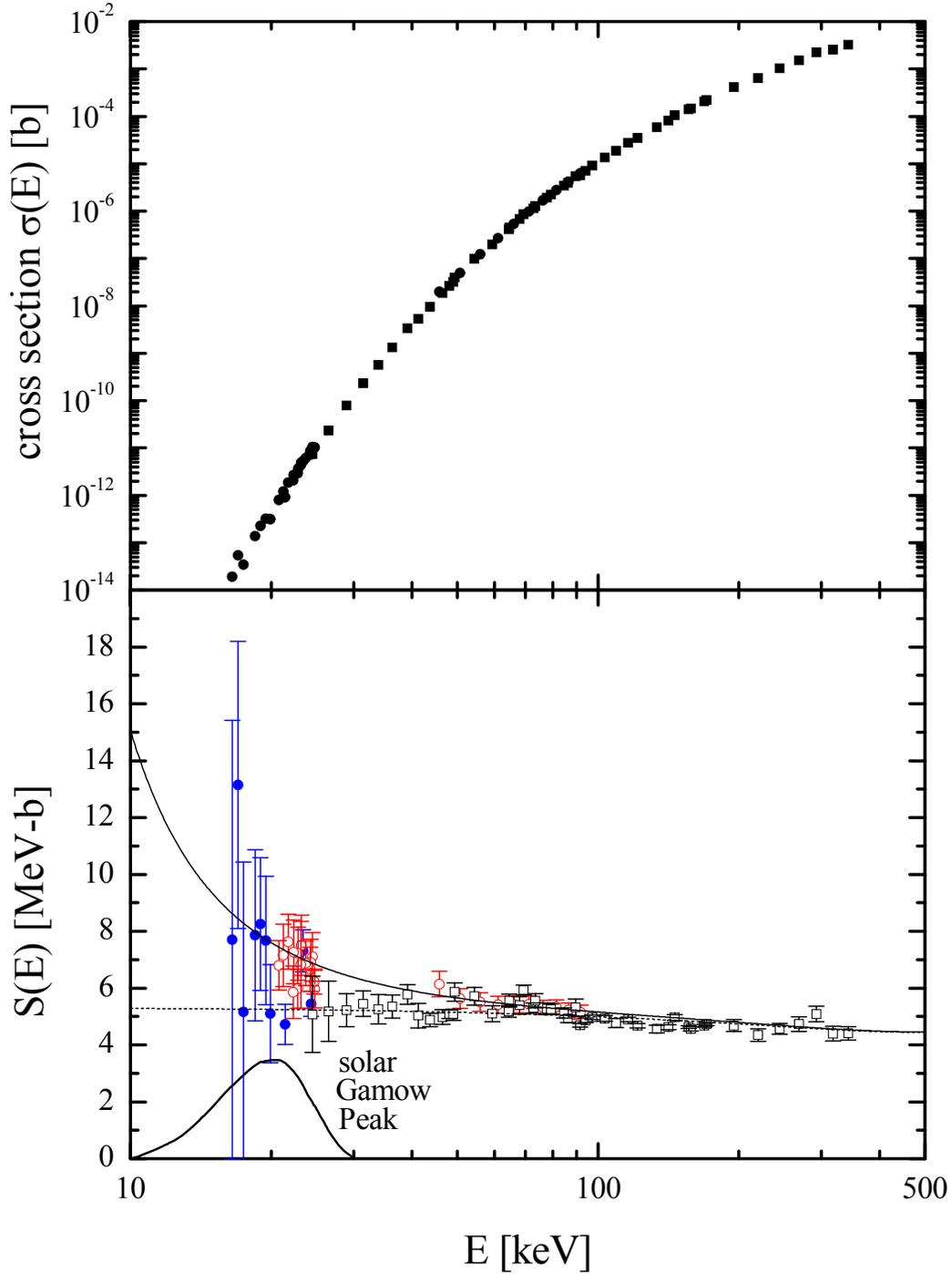

**Figure 1:** Energy dependence of the cross section $\sigma(E)$ and astrophysical S(E) factor for the $^3$He($^3$He,2p)$^4$He reaction. The Gamow peak shown is for solar conditions. The low-energy data have been obtained by LUNA I: red open circles with a detector telescope at Gran Sasso for E = 20 to 25 keV and at Bochum for E = 45 to 92 keV; blue filled circles with proton-proton coincidences at Gran Sasso. Also shown are the data of **[31]** (open squares). The increasing S(E) factor at low energies is due to the effects of electron screening (experiment $U_e$ = 294 ± 47 eV **[34]**, theory $U_{ad}$ = 220 eV **[38]**).

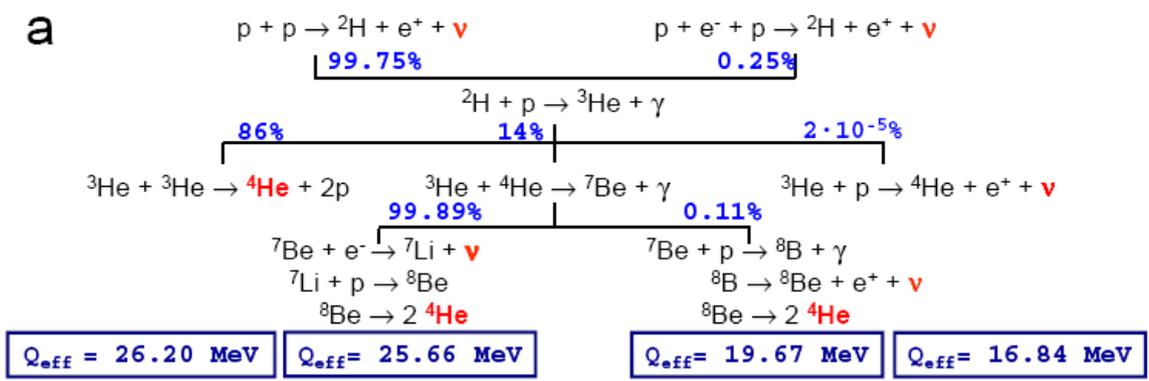
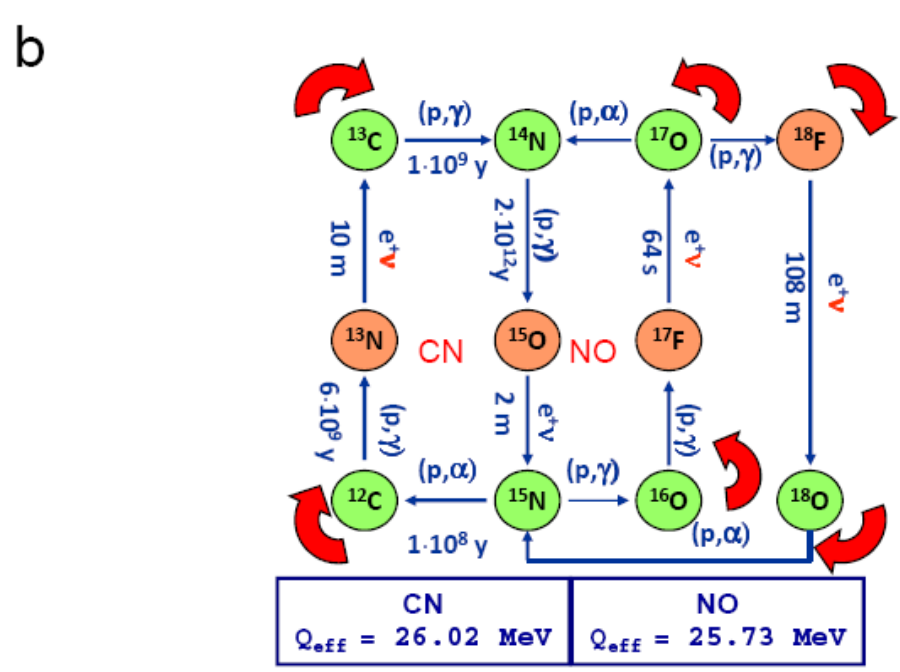

**Figure 2**: Reaction schemes of the pp-chain (a) and the CNO-cycle (b).

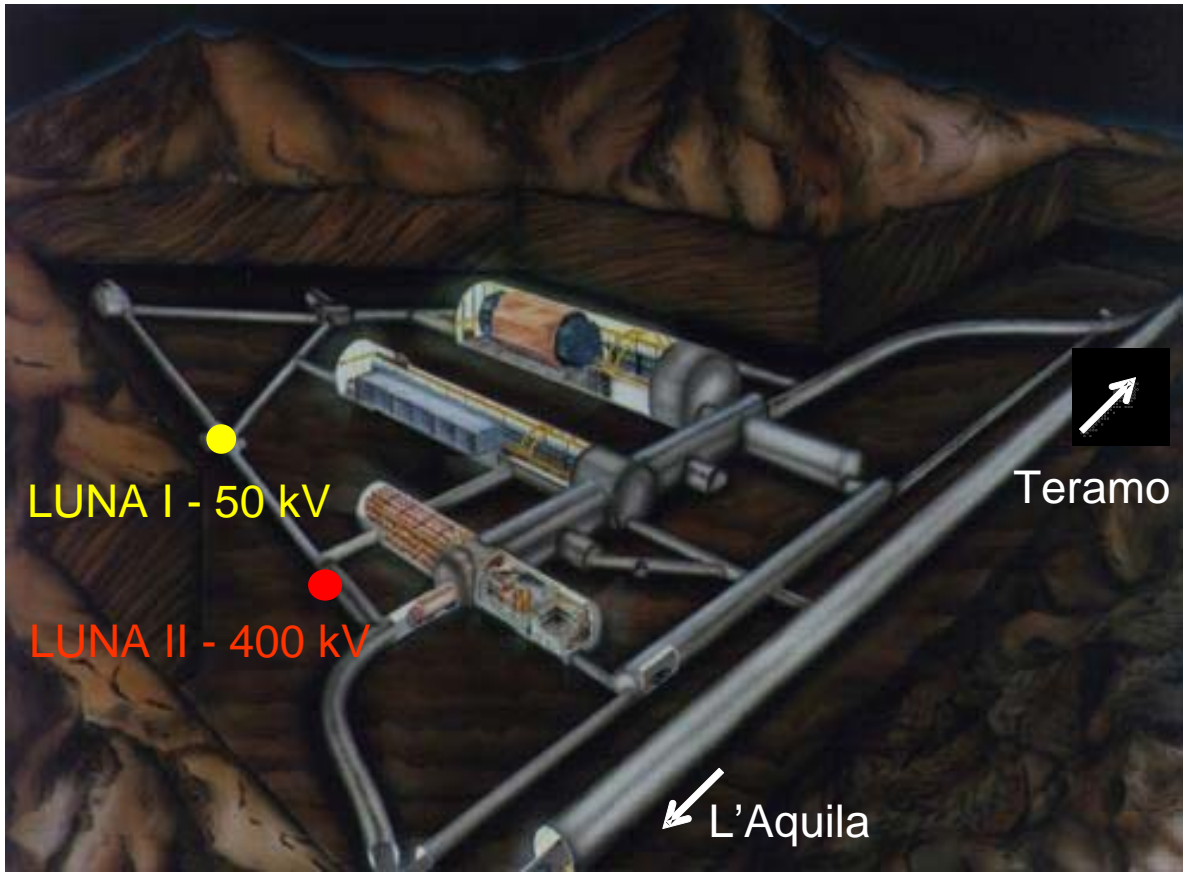

**Figure 3**: Floor plan of the Laboratori Nazionali del Gran Sasso (LNGS). The places allocated for the two LUNA accelerator facilities are marked.

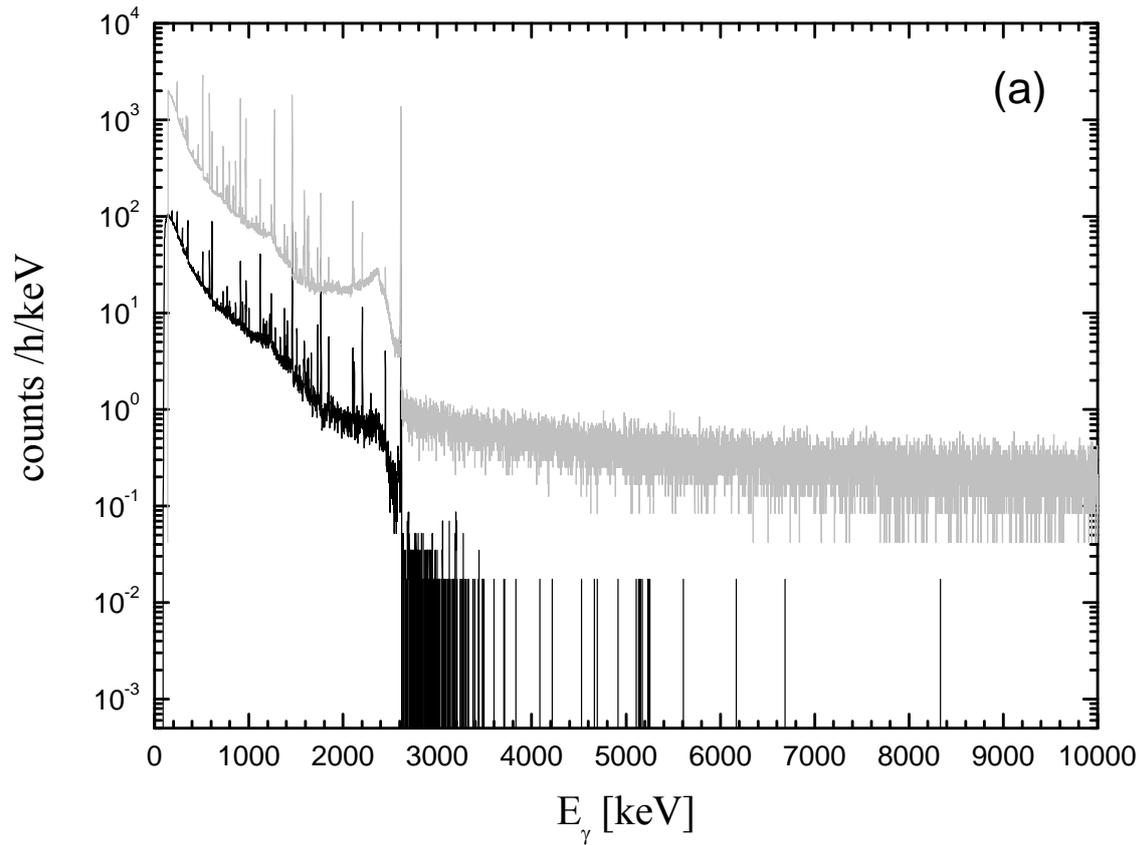

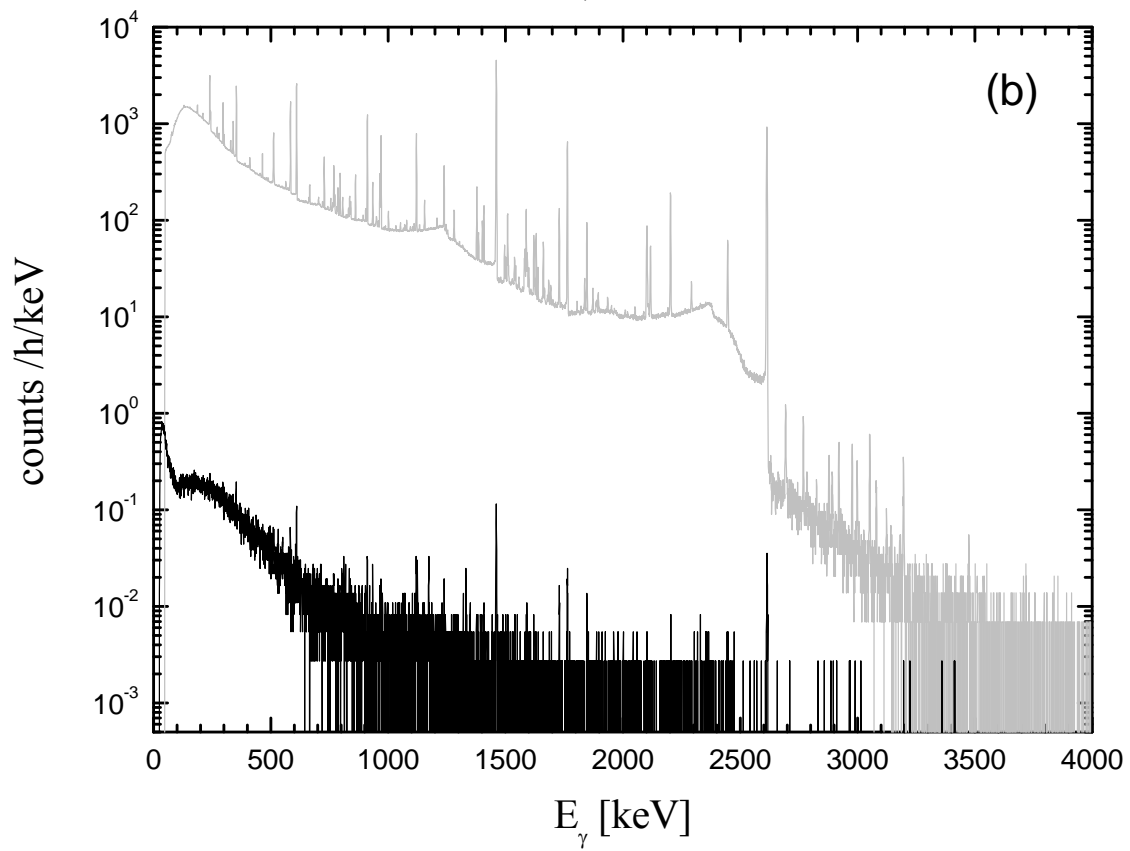

**Figure 4**: The upper panel illustrates spectra of γ-ray background as observed with a Ge detector placed outside (grey line) and inside (black line) of LNGS. The lower panel shows a comparison of low energy spectra inside LNGS without lead shielding (grey line) and including a heavy lead shield (black line).

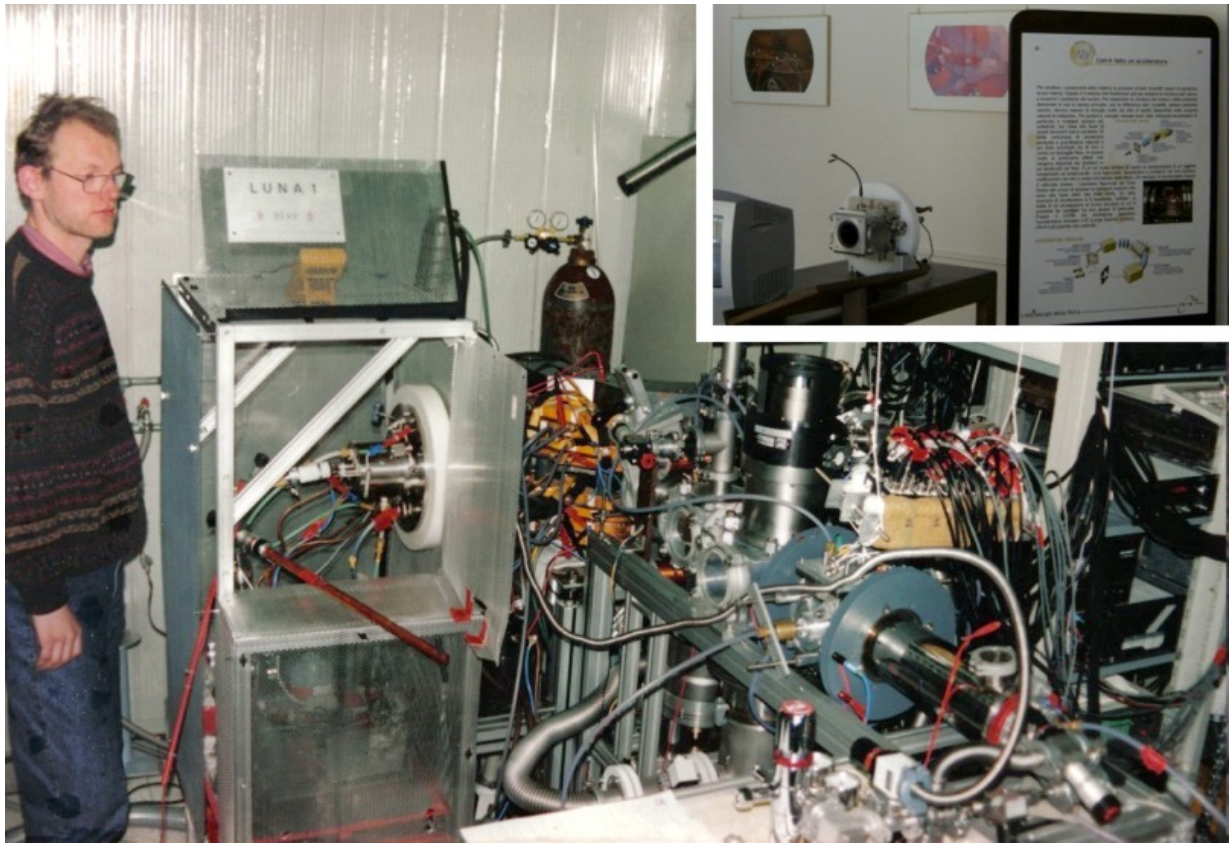

**Figure 5**: Photo of the LUNA I accelerator. On the left side one sees the ion source with the acceleration drift and on the right side the gas target with a beam calorimeter at the end. The inset (upper right corner) shows the accelerator part now displayed in a museum at Teramo, Italy.

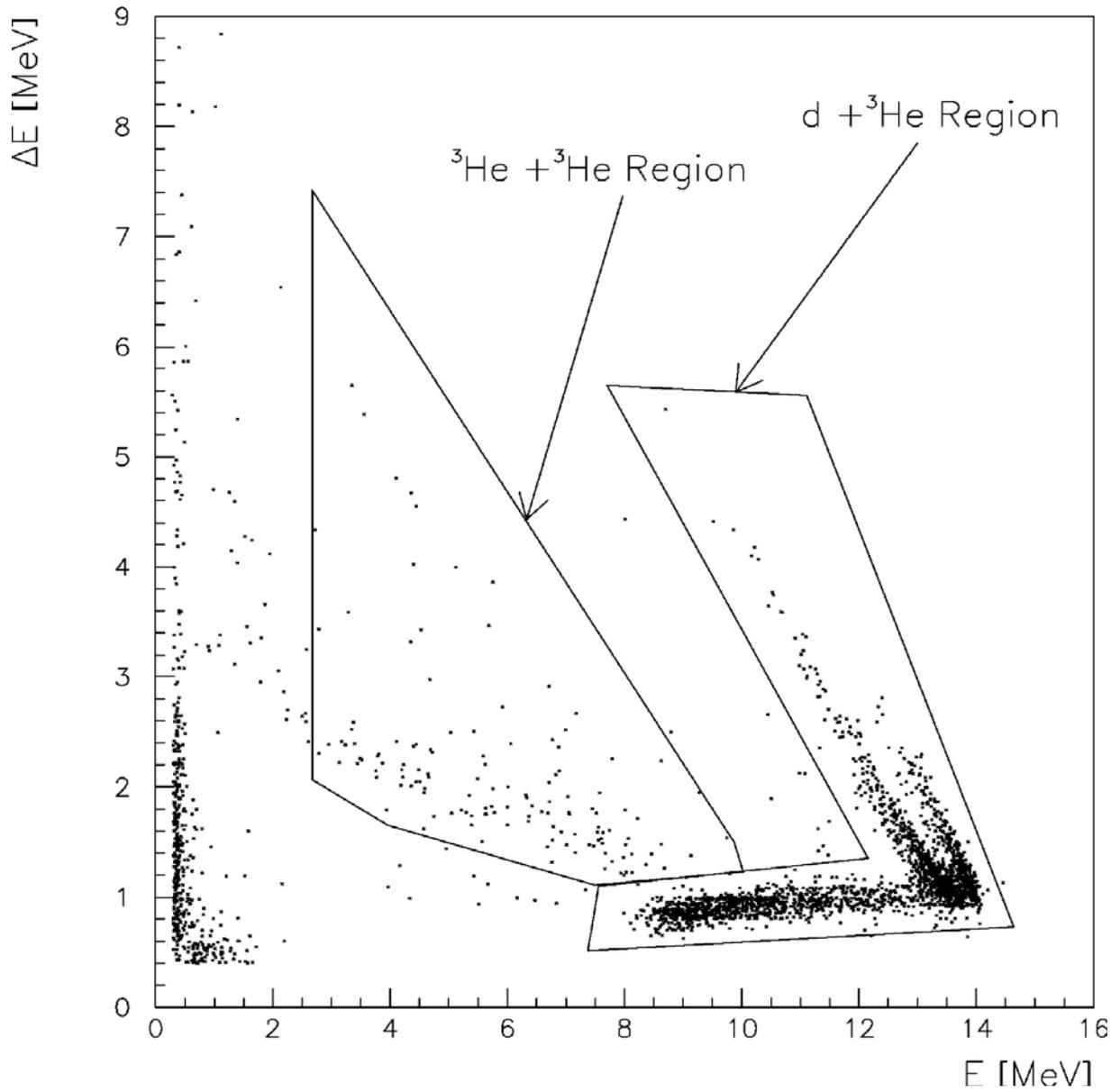

**Figure 6**: ΔE-E identification matrix of one telescope at E = 25 keV. The $^3$He+$^3$He and d+$^3$He selected regions are shown; note the beam-induced electronic noise at the left vertical edge of the matrix **[33]**.

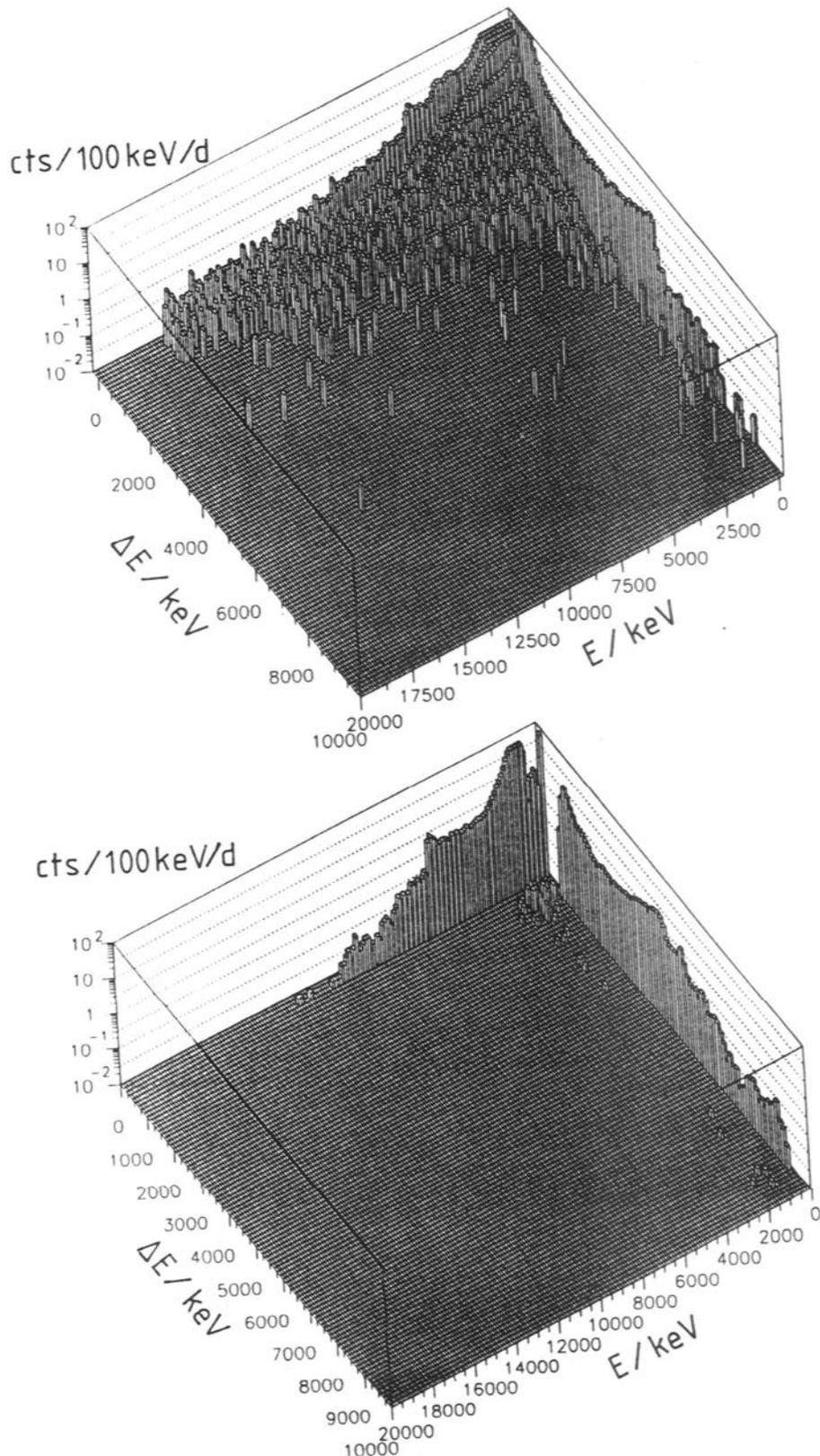

**Figure 7**: The upper panel shows a background run obtained in Bochum with a running time of 16 days, while the spectrum in the lower panel was measured at LNGS over 61 days. The reduction in background is clearly visible **[32]**.

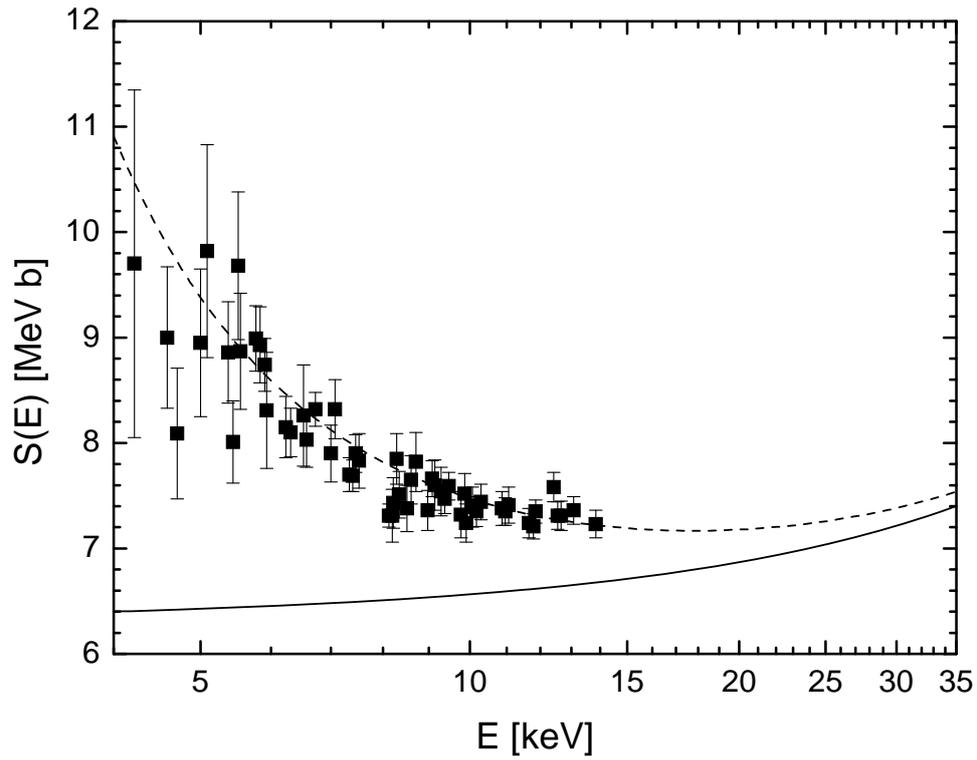

**Figure 8**: S factor data for the d($^3$He,p)$^4$He reaction obtained at the LUNA I facility **[41]**. The solid curve represents the S factor for bare nuclei and the dashed curve that for shielded nuclei with a screening potential $U_e$ = 132 eV. The bare S factor energy dependence was extrapolated from data sets at higher energy not shown here.

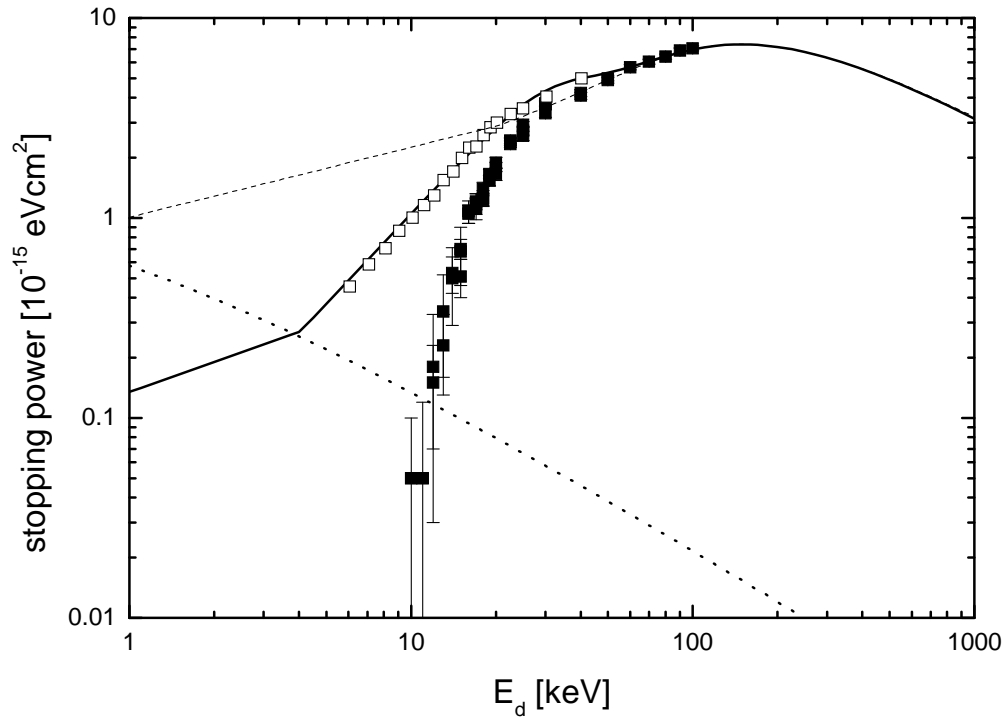

**Figure 9:** Total stopping-power data of deuterons in $^3$He gas at energies below the Bragg peak. The filled-in squares are from **[44]** and open squares from **[136]**. The dashed curve is the prediction of a compilation (SRIM-2000 **[42]**), based on data at energies near and above the Bragg peak, while in the recent version of the compilation low energy data were taken into account (solid line). The dotted curve represents the predicted nuclear stopping power **[42]**.

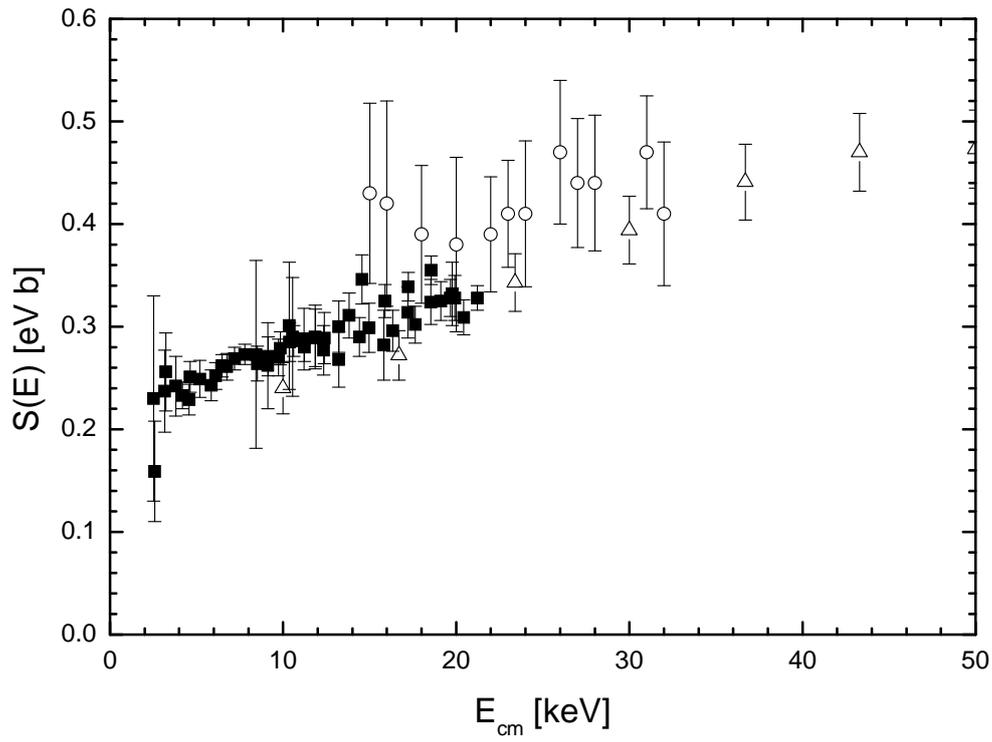

**Figure 10**: Results of the d(p,γ)³He reaction at LUNA: filled squares **[51]**; other data are open triangles **[130]** and open circles **[131]**.

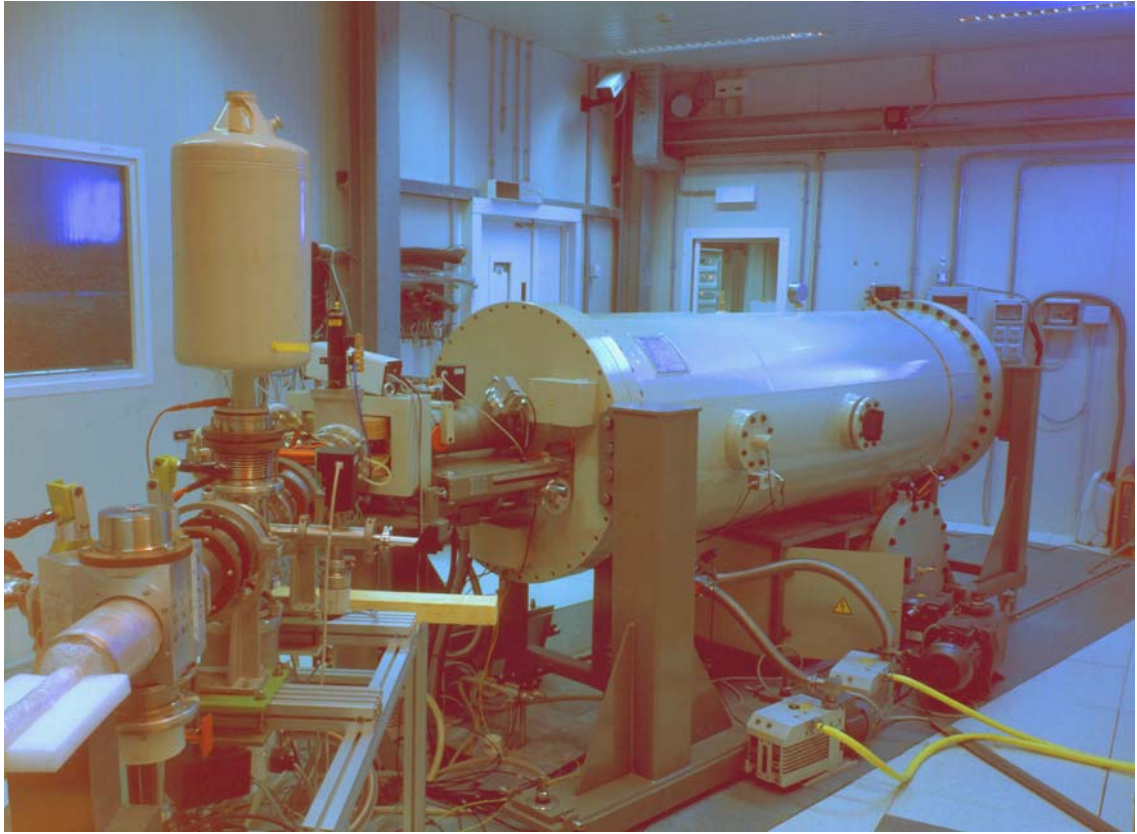

**Figure 11**: Photo of LUNA II during the installation phase.

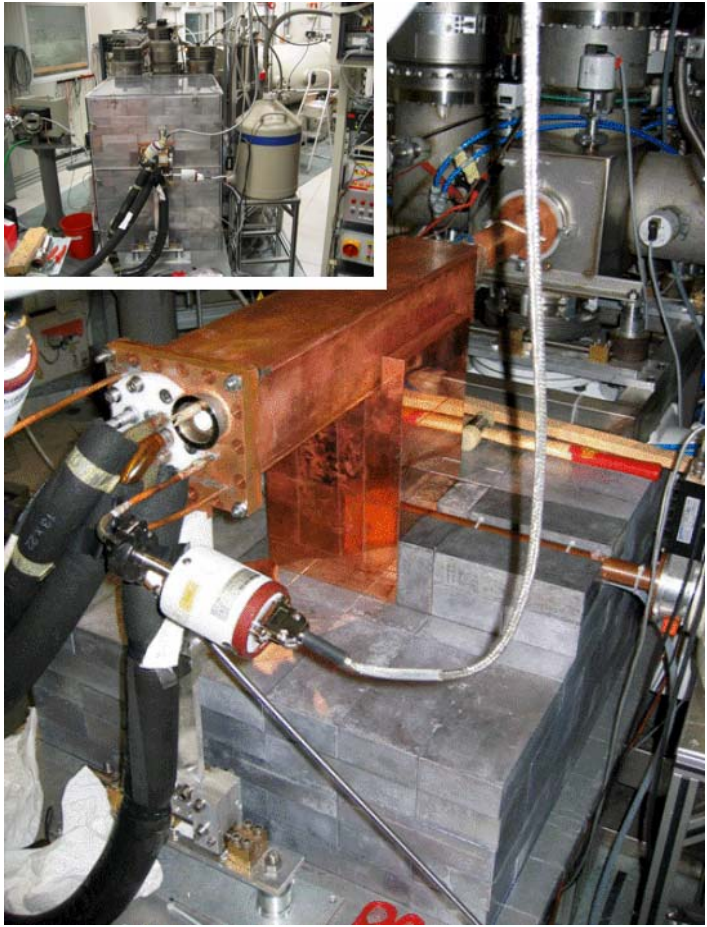

**Figure 12**: Photo of the construction phase of the lead shield for the $^3$He($\alpha,\gamma$)$^7$Be experiment. The copper enclosing the HPGe detector and the copper target chamber are visible. The inset shows the completed 0.3 m$^3$ lead shield surrounded by a Radon box.

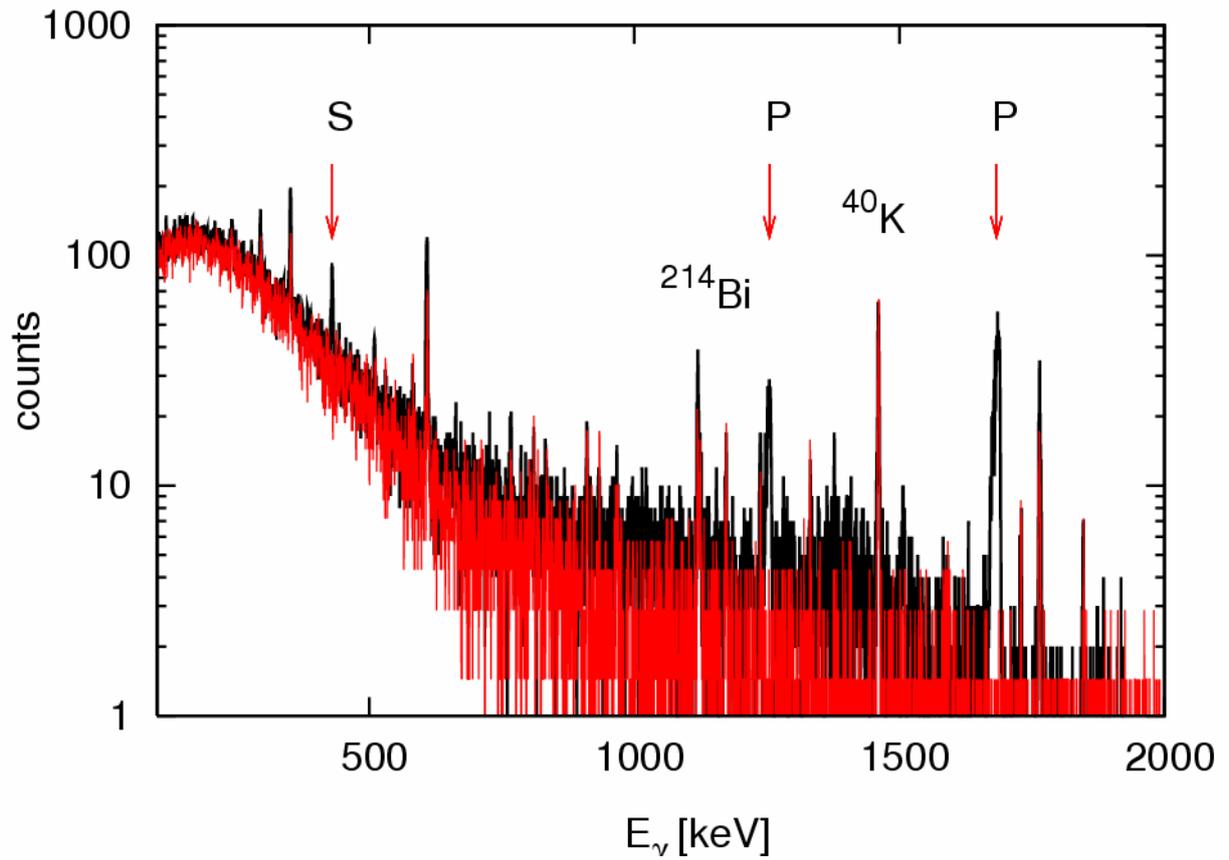

**Figure 13**: Gamma-ray spectrum at E = 93 keV (black solid line) compared with natural laboratory background (red line) normalized by time (31.2 days). Arrows indicate the primary transition peaks to the first excited state and to the ground state for the $^3$He($\alpha,\gamma$)$^7$Be as well as the secondary peak at $E_\gamma$ = 440 keV.

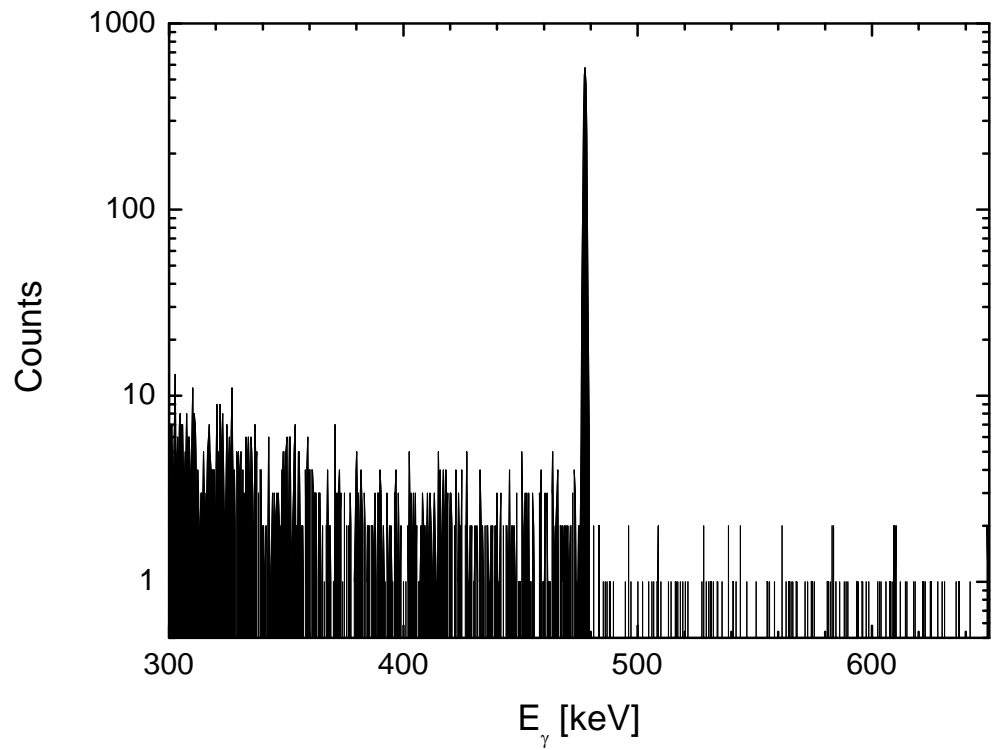

**Figure 14:** Gamma-ray spectrum of the $^7$Be activity for a sample irradiated at E = 148 keV. The spectrum was obtained with a HPGe detector of the LNGS Low-Level Laboratory. The total running time was 6 days and the measured activity about 480 mBq.

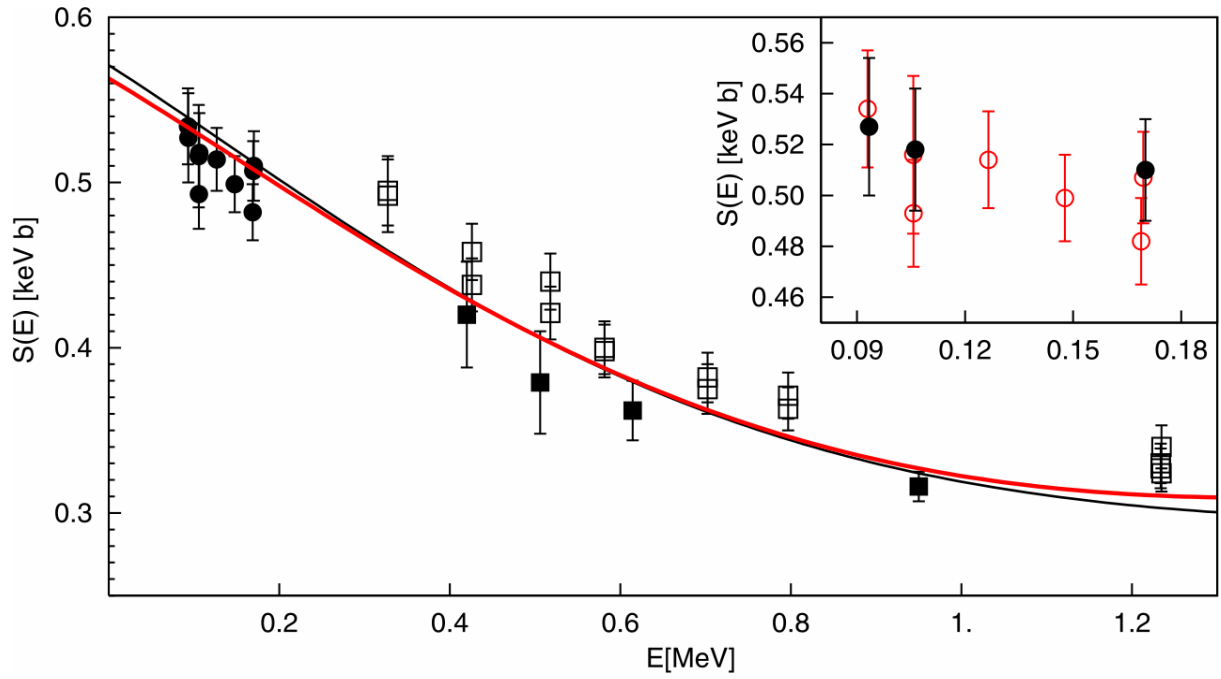

**Figure 15**: Astrophysical S factor for the $^3$He$(\alpha,\gamma)^7$Be reaction obtained from recent experiments. The filled squares are the data from **[66]**, the filled circles are the data from LUNA **[63, 64, 65],** and the open squares are the data from **[67]**. The black and red curves are the best fit to the data obtained re-scaling the S-factor curve as described in the text from **[66]** and **[67],** respectively. The inset shows a detailed comparison of the results from the prompt γ-ray measurement (filled circles) and the activity method (open circles).

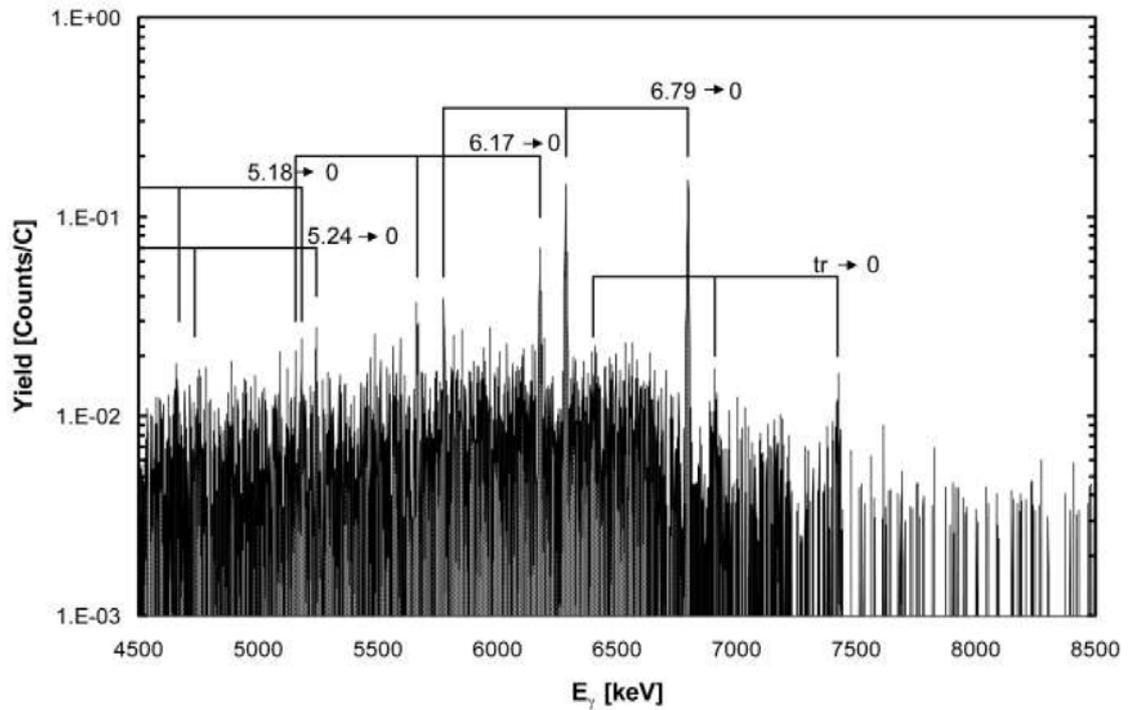

**Figure 16**: Gamma-ray spectrum of the $^{14}N(p,\gamma)^{15}O$ reaction obtained at E = 130 keV over an accumulated charge of 210 C **[73]**. The positions of the ground state transition and the secondary transitions are indicated.

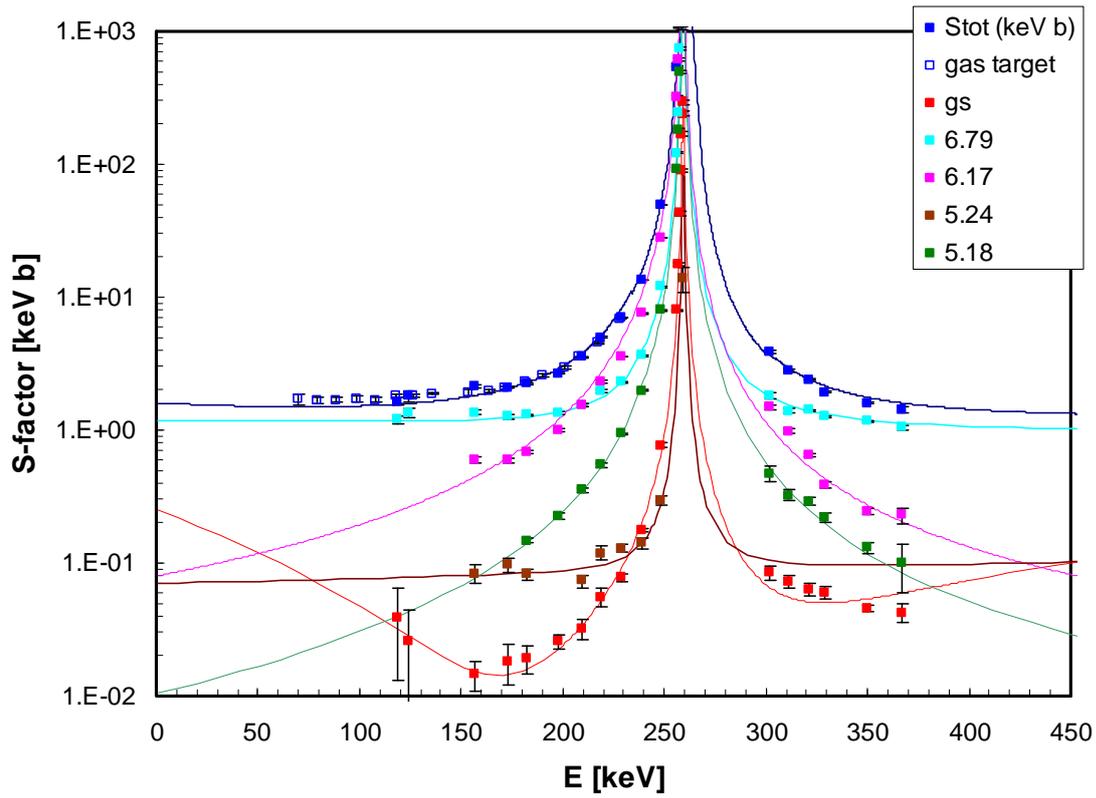

**Figure 17**: Astrophysical S factor of $^{14}N(p,\gamma)^{15}O$ obtained at LUNA with the high resolution setup (HPGe + solid target **[72, 73]**), where each transition has been detected (filled symbol) and the high efficiency setup (BGO + gas target **[74, 75]**), where the summing peak has been detected (open symbol). The solid colored lines represent the R-matrix calculations from **[73]**.

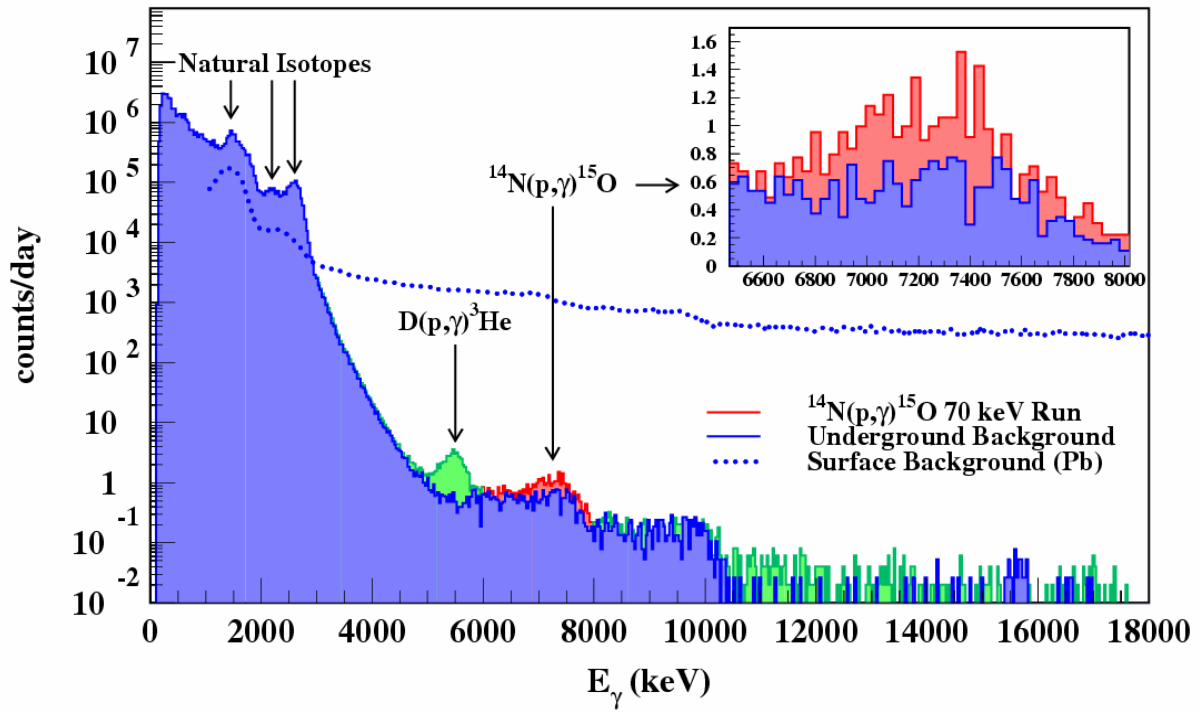

**Figure 18**: Gamma-ray spectrum obtained with the 4π BGO detector at E = 70 keV. The running time is 49 days and the accumulated charge is 928 C. The natural background contribution is shown in blue and the beam induced background in green. The region of interest for the $^{14}N(p,\gamma)^{15}O$ reaction is blown up in the inset.

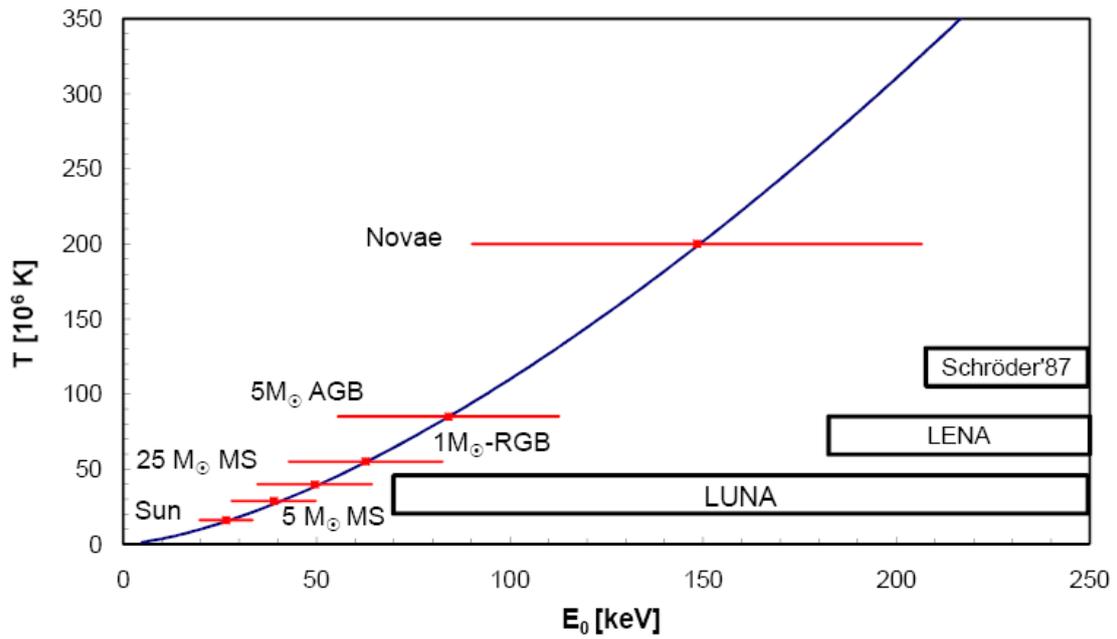

**Figure 19:** The $^{14}N(p,\gamma)^{15}O$ Gamow energy $E_0$ for different astrophysical sites: sun, 5 $M_\odot$ and 25 $M_\odot$ Main Sequence, 1 $M_\odot$ Red Giant Branch (RGB), 5 $M_\odot$ Asymptotic Giant Branch (AGB) stars, and Novae; their respective central temperature are shown as ordinate. The LUNA results reach Novae and AGB stars. The energy regions covered by the experiments of Runkle et al. (LENA) **[77]** and Schröder et al. **[78],** cover only partially Novae.

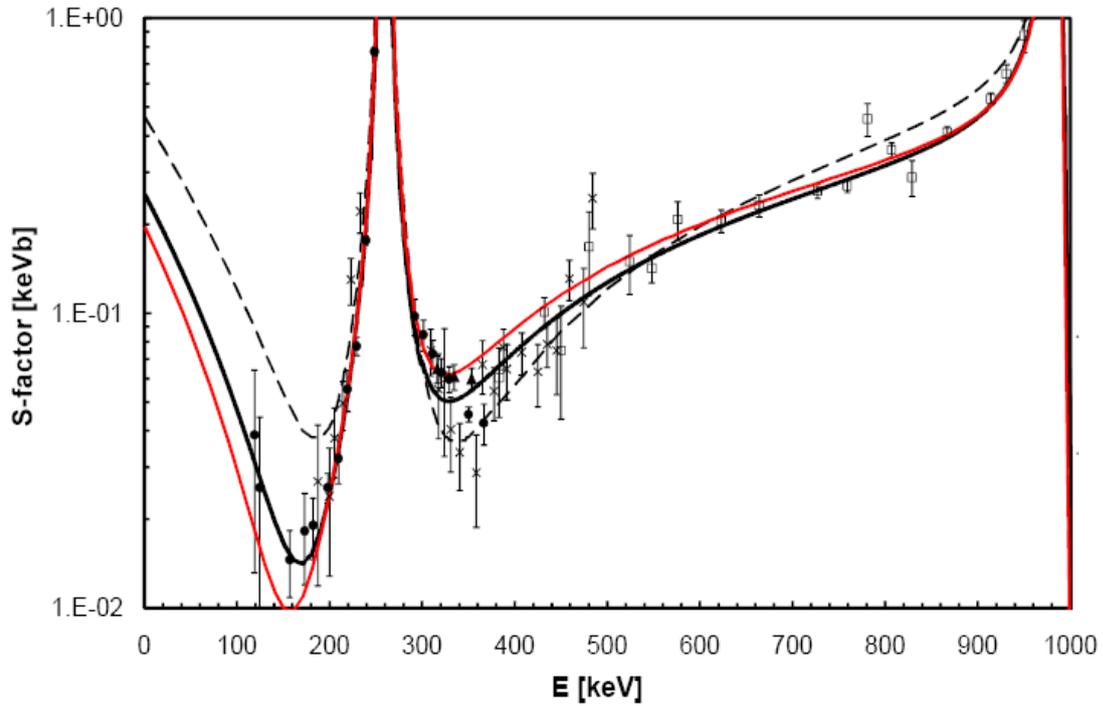

**Figure 20**: Astrophysical S factor for the ground state transition, filled circle are data from **[73]**, filled triangle **[79]**, crosses **[77]** and open square **[78]**. The black solid line is the R-matrix fit done in **[73]**, dashed line **[77]** and the red is the R-matrix fit considering only the data obtained in experiment done with small summing, i.e. **[79]** and in part **[78]**.

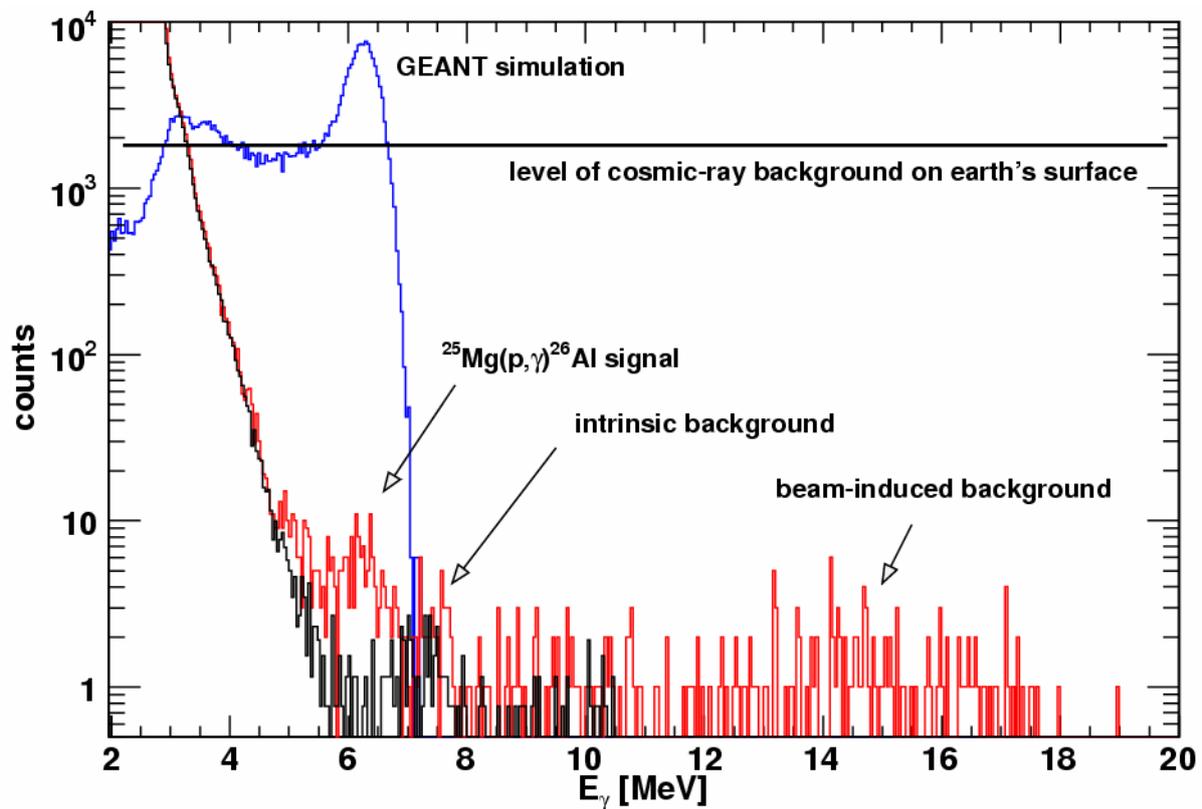

**Figure 21**: Gamma-ray spectrum of the $^{25}$Mg(p,γ)$^{26}$Al resonance at E = 93 keV obtained with the BGO detector at LNGS and associated background. A high statistics GEANT4 simulation [132] is shown to demonstrate the correct γ-ray energy of the peak.

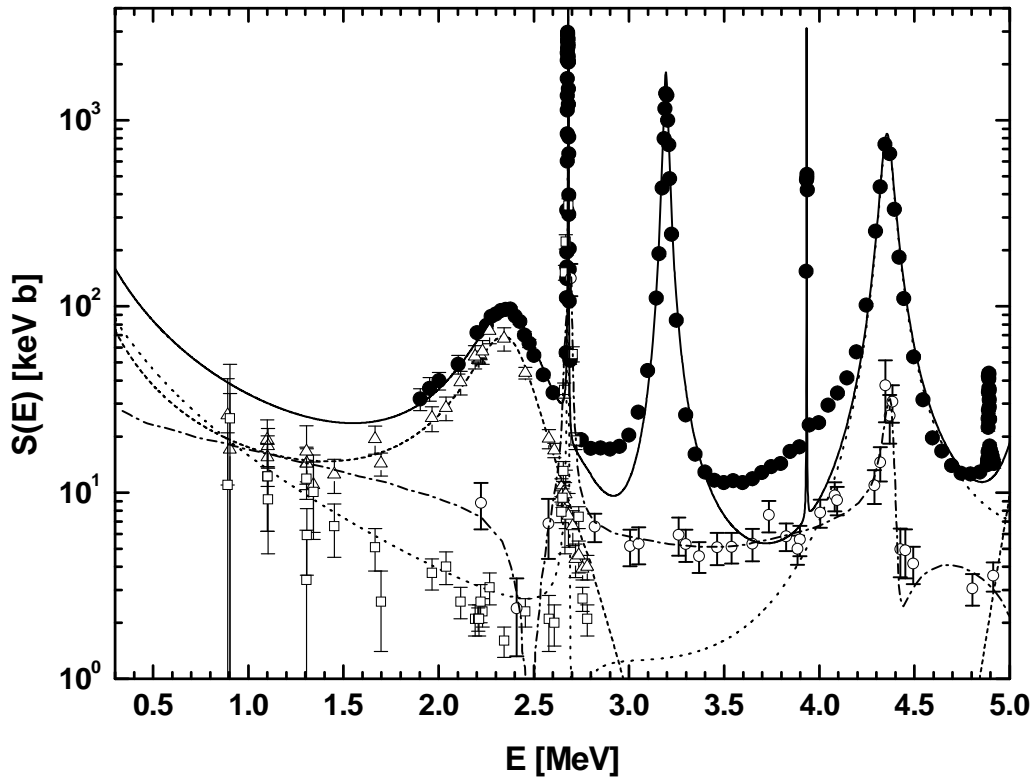

**Figure 22:** The present status of the S factor data for $^{12}C(\alpha,\gamma)^{16}O$. The total S factor measurement of ERNA (filled-in circles) **[12]** is compared with recent E1 (open triangles) and E2 (open squares) γ-ray measurements **[98]** and the $E_x$ = 6.05 MeV cascade data (open circles) **[100]**. The solid line represents the sum of the single amplitudes of an R matrix fit **[99]** (the dotted and dashed lines are the E1 and E2 amplitudes, respectively). In addition, the R matrix fit of **[100]** to their cascade data (dotted-dashed line) is shown. The latter component is not included in the sum and might explain the high yield in the ERNA data between the resonances.

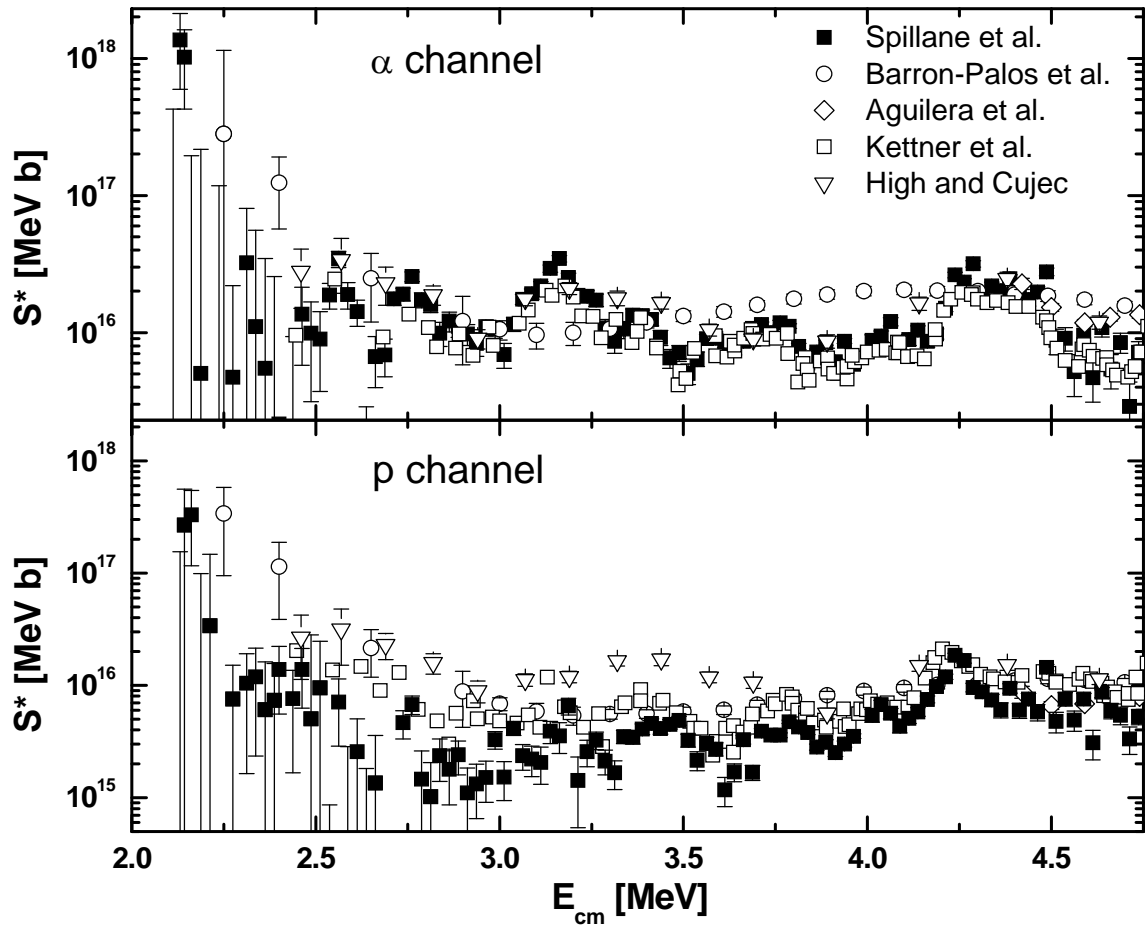

**Figure 23:** Modified astrophysical S(E)* factor of the fusion process $^{12}C+^{12}C$ for the α and proton channels **[109]**.

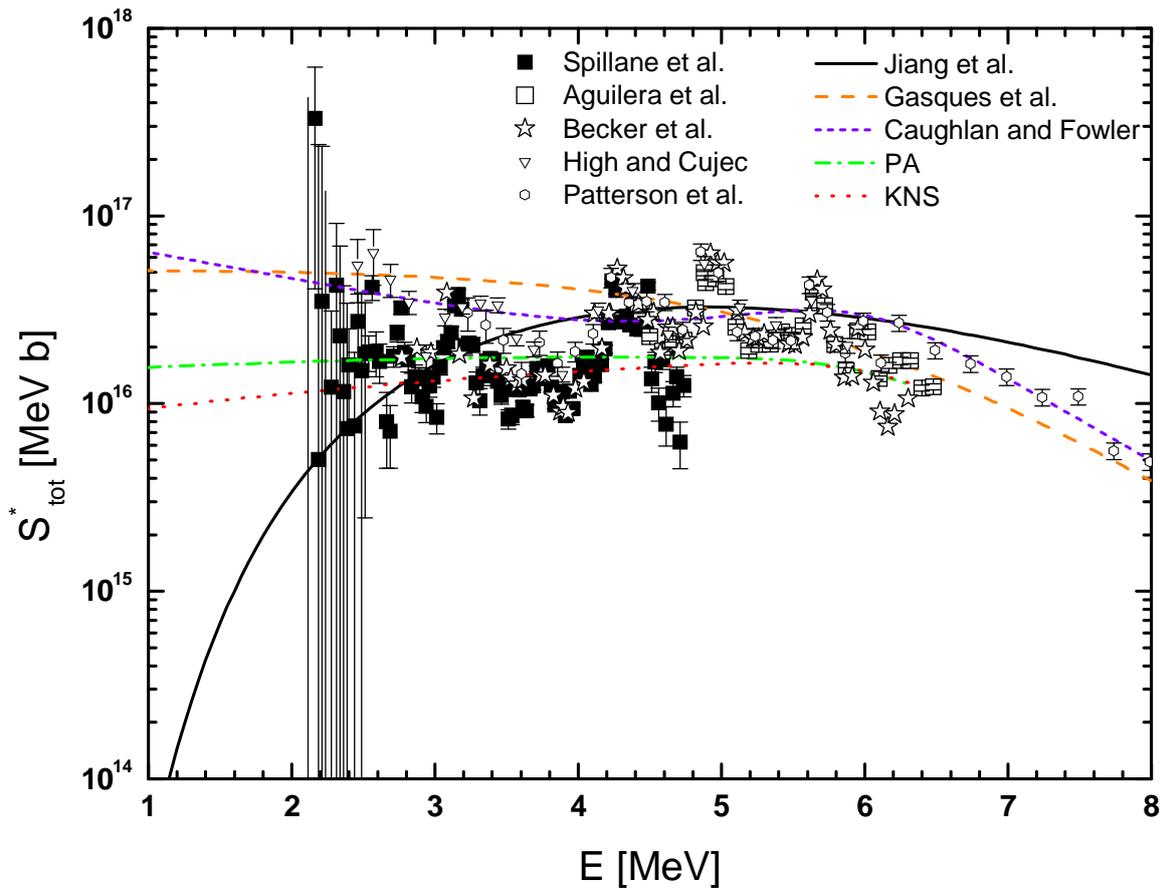

**Figure 24**: Comparison of the experimental data **[109, 111, 115, 132, 133]** for the total $S^*_{tot}$ factor with various theoretical calculations **[110, 111, 112, 114]**. The resonance structures exist over the full energy range and are most likely superimposed on a flat non-resonant contribution.

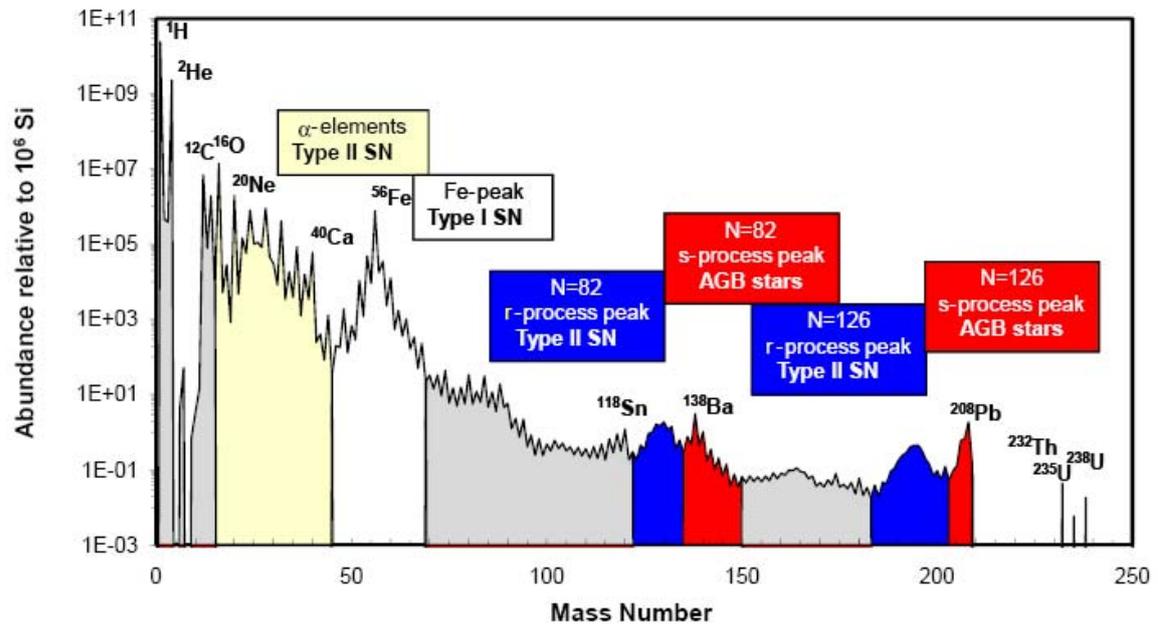

**Figure 25**: Solar abundance **[134]** and the production sites for the trans-iron elements.

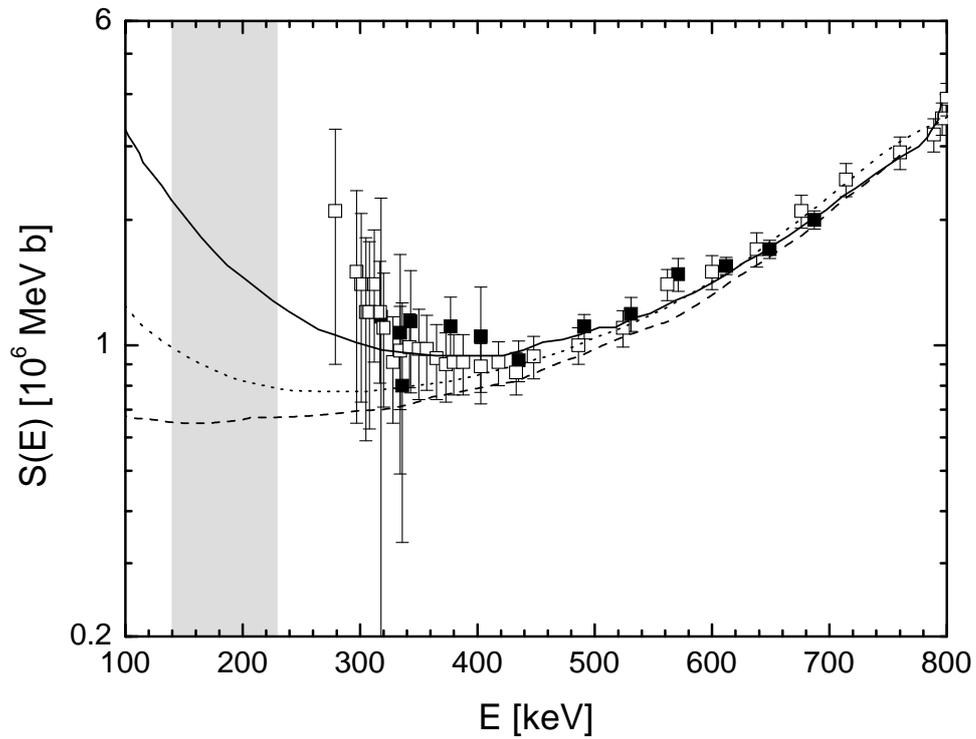

**Figure 26**: Astrophysical S factor of the $^{13}C(\alpha,n)^{16}O$ reaction from the recent work of **[135]** (filled-in squares) and previous data **[120]**. The different R-matrix extrapolations are also displayed: solid line **[135]** (almost identical with **[123]**), dotted line **[122]**, and dashed line **[121]**. The Gamow window for the s-process in $^{13}C$ pockets of AGB stars is given by the shaded area.

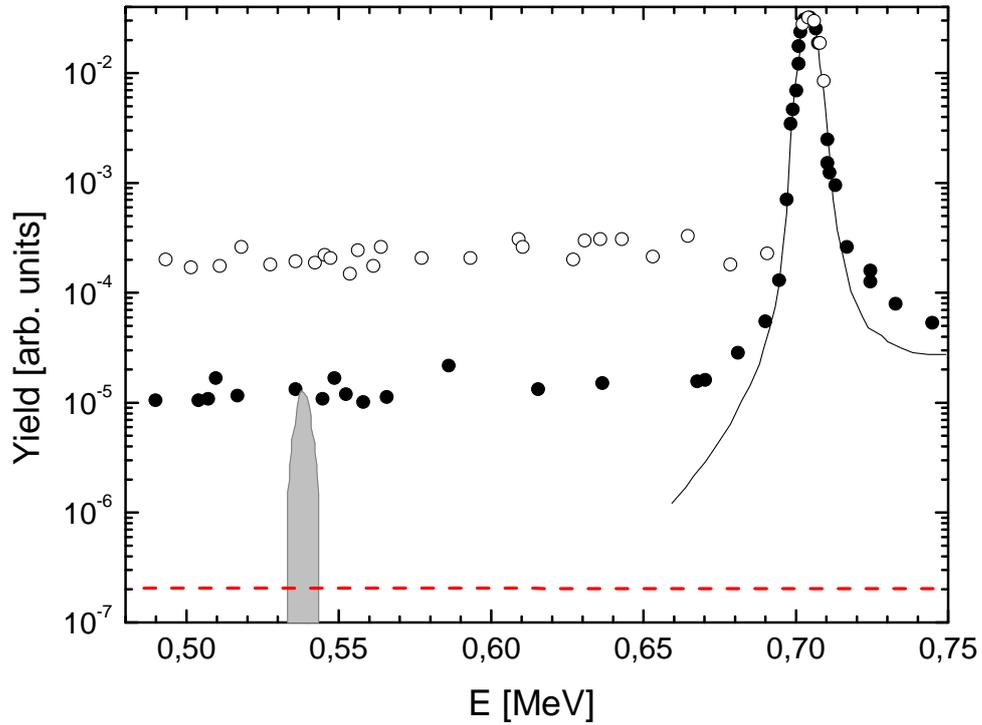

**Figure 27**: Low-energy region of the excitation function of the $^{22}$Ne($\alpha$,n)$^{25}$Mg reaction from the work of **[127]** (filled-in circles) and a previous measurement **[120]** (open circles). All data points below E = 0.68 MeV are only upper limits. An expected resonance at E = 0.537 MeV is shown as the shaded area with an upper limit for its strength of $\omega\gamma$ < 60 neV **[127]**. The dashed line represents a conservative sensitivity limit in a future underground experiment.